\documentclass{svjour3}

\usepackage{verbatim}
\usepackage{color}		
\usepackage{epsfig}
\usepackage{graphicx}
\usepackage{amsfonts}
\usepackage{amssymb}
\usepackage{ifpdf}
\usepackage{epstopdf}
\usepackage{amssymb}
\usepackage[all]{xy}
\usepackage{times}

\usepackage[compatibility=false]{caption}
\usepackage{subcaption}

\newcommand{\ms}[1]{\ifmmode%
\mathord{\mathcode`-="702D\it #1\mathcode`\-="2200}\else%
$\mathord{\mathcode`-="702D\it #1\mathcode`\-="2200}$\fi}

\newcommand{\xb}{x_{\rm b}}
\newcommand{\xd}{x_{\rm d}}
\newcommand{\xu}{x_{\rm u}}
\newcommand{\yb}{y_{\rm b}}
\newcommand{\yd}{y_{\rm d}}
\newcommand{\yu}{y_{\rm u}}

\newcommand{\smallbox}{{\scriptscriptstyle{\boxtimes}}}

\begin{document}


\title{Flow-based reputation with uncertainty}
\subtitle{Evidence-Based Subjective Logic}

\author{}


\institute{ }


\maketitle



\vskip-35mm

\noindent 
{Boris \v{S}kori\'{c}, Sebastiaan J.A. de Hoogh, and Nicola~Zannone}

\vskip10mm

\begin{abstract}
The concept of
reputation is widely used 
as a measure of trustworthiness  based on ratings from members in a community.
The adoption of reputation systems, however, relies on their ability to capture the actual trustworthiness of a target.
Several reputation models for aggregating trust information
have been proposed in the literature.
The choice of model has an impact on the reliability of the aggregated trust information 
as well as on the procedure used to compute reputations.
Two prominent models are flow-based reputation (e.g., EigenTrust, PageRank) and Subjective Logic
based reputation.
Flow-based models provide an automated method to aggregate trust information, 
but they are not able to express the level of uncertainty in the information.
In contrast, Subjective Logic extends probabilistic models with an explicit notion of uncertainty, but the calculation of reputation depends on the structure of the trust network and often requires information to be discarded. These are severe drawbacks.

In this work, we observe that the `opinion discounting' operation in Subjective Logic has a number of basic problems.
We resolve these problems by providing a new discounting operator that describes the flow of evidence from one party to another.
The adoption of our discounting rule results in a consistent
Subjective Logic algebra that is entirely based on the handling of evidence.
We show that the new algebra enables the construction
of an automated reputation assessment procedure for arbitrary trust networks,
where the calculation no longer depends on the structure of the network, and
does not need to throw away any information.
Thus, we obtain the best of both worlds: flow-based reputation and consistent handling of uncertainties.

\keywords{Reputation systems \and Evidence Theory \and Subjective Logic \and Flow-based reputation models}
\end{abstract}

\section{Introduction}

Advances in ICT and the increasing use of the Internet have resulted in changes in the way people do everyday things and interact with each other.
Everything people do happens online, significantly increasing the number of business transactions carried out daily over the Internet. 
Often, users have to decide whether to interact with services or users with whom they have never interacted before.
Uncertainty about services and users' behavior is often perceived as a risk \cite{AsnarZ08} and, thus, it can restrain a user from engaging in a transaction with unknown parties.
Therefore, to fully exploit the potential of online services, platforms and ultimately online communities, it is necessary to establish and manage trust amongst the parties involved in a transaction \cite{Grandison:2000,TrivellatoZE14,Xiong:2004}.

Reputation is widely adopted to build trust among users in online communities where users do not know each other beforehand.
The basic idea underlying reputation is that a user's past experience as well as the experience of other users influences his decision whether to repeat this interaction in the future. 
Thus, reputation provides an indication of services' and users' trustworthiness based on their past behavior \cite{Smeltzer1997}.
When a user has to decide whether to interact with another party, he can consider its reputation and start the transaction only if it is trustworthy. 
Therefore, a reputation system, which helps managing reputations (e.g., by collecting, distributing and aggregating feedback about services and users' behavior), becomes a fundamental component of the trust and security architecture of any online service or platform \cite{Vavilis2014}.

The application and adoption of reputation systems, however, relies on their ability to capture the actual trustworthiness of the parties involved in a transaction \cite{impactict}.
The quality of a reputation value depends on the amount of information used for its computation \cite{Govindan12,Josang12,Liu2012438}.
A reputation system should use ``sufficient'' information.
However, it is difficult to establish the minimum amount of information required to compute reputation; 
also, different users may have a different perception based on their risk attitude~\cite{AsnarZ08}.
For instance, some users may accept to interact with a party which has a high reputation based on very few past transactions, while other users might require more evidence of good behavior.
Therefore, a reputation system 
should provide a level of confidence in the computed reputation, for instance based on the amount of information used in the computation \cite{josang2001logic,Ries11}.
This additional information will provide deeper insights to users, helping them decide whether to engage in a transaction or not.
In addition, reputation systems should provide an effective and preferably automated method to aggregate the available trust information
and compute reputations from it.

Reputation systems usually rely on a mathematical model to aggregate trust information and compute reputation \cite{Hoffman:2009}. 
Several mathematical models for reputation have been proposed in the literature.
These models can be classified with respect to the used mathematical foundations, e.g.\ summation and averaging \cite{huynh2006integrated}, probabilistic models \cite{josang2001logic,josang2007dirichlet,Muller2013}, flow-based \cite{pagerank,kamvar2003eigentrust,salsa,flowbase}, fuzzy metrics \cite{bharadwaj2009fuzzy,song2005trusted}.
As pointed out in \cite{Vavilis2014}, the choice of the type of model has an impact on the type and amount of trust information as well as on the procedure used to compute reputation.

Among the others, two prominent reputation models are the flow-based model and Subjective Logic (SL) \cite{josang2001logic}. 
Flow-based reputation models use Markov chains as the mathematical foundation.
Flow-based models provide an automated method to aggregate all available trust information. 
However, they are not able to express the level of confidence in the obtained reputation values.
On the other hand, SL is rooted in the well-known Dempster-Shafer theory \cite{Dempster-Shafer}.
SL provides a mathematical foundation to deal with opinions and has the natural ability to express uncertainty explicitly. 
Intuitively, uncertainty incorporates a margin of error into reputation calculation due to the (limited) amount of available trust information.
SL uses a \emph{consensus} operator `$\oplus$'
to fuse independent opinions and a \emph{discounting} operator `$\otimes$'
to compute trust transitivity.
This makes SL a suitable mathematical framework for handling trust relations and reputation, especially when limited evidence is available. 
However, the consensus operator is rooted in the theory of evidence, while the discounting operator is based on a probabilistic interpretation of opinions.
The different nature of these operators leads to a lack of ``cooperation'' between them. 
As a consequence, the calculation of reputation depends on the shape of the \emph{trust network}, the graph
of interactions, in which nodes represent the entities in the system and edges are labeled with opinions.
Depending on the structure of the trust network, some trust information may have to be discarded to 
enable SL-based computations.

Our desideratum is to have a reputation system which has the advantages of both flow-based reputation models and SL.
In particular, the goal of this work is to devise the mathematical foundation for a flow-based reputation model with uncertainty.
We make the following contributions towards this goal:
\begin{itemize}
\item
We observe that the discounting rule `$\otimes$' in SL does not have a natural
interpretation in terms of evidence handling.
We give examples of counterintuitive behavior of the $\otimes$ operation.
\item
We present a brief inventory of the problems that occur when one tries to combine
SL with flow-based reputation metrics.


\item
We present a simplified justification of the mapping between evidence and opinions in~SL.
\item
We introduce a new scalar multiplication operation in SL
which corresponds to a multiplication of evidence.
Our scalar multiplication is consistent with the consensus operation 
(which amounts to addition of evidence), and
hence satisfies a distribution law, namely
$\alpha\cdot(x\oplus y)=(\alpha\cdot x)\oplus(\alpha\cdot y)$.
\item
We introduce a new discounting rule $\boxtimes$.
It represents the \emph{flow of evidence} from one party to another.
During this flow, lack of trust in the party from whom evidence is received
is translated into a reduction of the amount of evidence; this reduction
is implemented using our new scalar multiplication rule.
Our new discounting rule satisfies 
$x\boxtimes(y\oplus z)=(x\boxtimes y)\oplus(x\boxtimes z)$.
This right-distribution property resolves one of the problems of SL.
\end{itemize}
Our new discounting rule multiplies evidence instead of opinions and is thus 
fully based on the handling of evidence. 
In contrast to the old discounting rule, the new one does not have associativity.
This, however, does not pose a problem since the flow of evidence has a well defined direction.
The adoption of our discounting rule results in an opinion algebra that is entirely centered on the handling of evidence.
We refer to it as {\bf Evidence-Based Subjective Logic} (EBSL).

We show that EBSL provides  a solid foundation for the development of reputation models able to 
express the level of confidence in computed reputations.
Moreover, having an opinion algebra rooted in a single foundation makes it possible to define an automated procedure to compute reputation for arbitrary trust networks.
We believe that having an automated procedure is a critical and necessary condition for the implementation and adoption of reputation systems in online services.
To demonstrate the applicability of EBSL to the development of reputation systems, we make the following contributions:
\begin{itemize}
\item
We show that replacing SL's discounting operation $\otimes$ by the new $\boxtimes$
solves all the problems that usually occur when one tries to combines flow-based reputation with SL,
in particular the problem of evidence double-counting and the ensuing necessity to make
computations graph-dependent and to discard information.
\item 
Using EBSL, we construct a simple iterative algorithm that computes 
reputation {\em for arbitrary trust networks} without discarding any information.
Thus, we achieve our desideratum of automated computation of flow-based reputation with
uncertainty.
\end{itemize}
We stress that this is only one out of many possible reputation models that can be constructed on top of EBSL.
EBSL can be used as a foundation for other existing reputation models, e.g.\ 
reputation models based on random walks \cite{GkorouVPE13}.
The investigation of these models is an interesting direction for future work.


The remainder of the paper is organized as follows.
The next section presents an overview of flow-based reputation models and SL.
Section~\ref{sec:flowSL} discusses the limitations of SL, and Section~\ref{sec:combining-FB-SL} illustrates these limitations when combining flow-based reputation and SL.
Section~\ref{sec:SLrevisited} revisits SL and introduces the new EBSL scalar multiplication and discounting operators.
Section~\ref{sec:SLflow} presents our flow-based reputation model with uncertainty along with an iterative algorithm that computes reputation for arbitrary trust networks. 
Section~\ref{sec:experiments} presents an evaluation of  our approach  using both synthetic and real data.
Finally, Section~\ref{sec:related} discusses related work, and Section~\ref{sec:conclusions} concludes the paper providing directions for future work.

\section{Preliminaries}\label{sec:preliminaries}
In this section we present an overview of flow-based reputation models (based on \cite{flowbase}) and of Subjective Logic (based on \cite{josang2001logic}).
We also introduce the notation used in the remainder of the paper.

\subsection{Flow-based reputation}

Flow-based reputation systems \cite{pagerank,kamvar2003eigentrust,salsa,flowbase} are based on the notion of transitive trust.
Intuitively, if an entity $i$ trusts an entity $j$, it would also have some trust in the entities trusted by $j$. 

Typically, each time user $i$ has a transaction with another user $j$, she may rate the transaction as positive, neutral, or negative.
In a flow-based reputation model, these ratings are aggregated in order to obtain a Markov chain. 
The reputation vector (i.e., the vector containing all reputation values) is computed as the steady state vector of the Markov chain; one starts with a vector of initial reputation values and then repeatedly applies the Markov step until a stable state has been reached.
This corresponds to taking more and more indirect evidence into account.


Below, we present the metric proposed in \cite{flowbase} (with slightly modified notation)
as an example of a flow-based reputation system.

\begin{example}
Let $A$ be a matrix containing aggregated ratings for $n$ users.
It has zero diagonal (i.e., $A_{ii}=0$).
The matrix element $A_{ij}$  (with $i\neq j$) represents the aggregated ratings given to $j$ by~$i$.
Let ${\bf s}\in [0, 1]^n$, with ${\bf s} \neq {\bf 0}$, be a `starting vector' containing 
starting values assigned to all users. 
Let $\alpha \in [0, 1]$ be a weight parameter for the importance of indirect vs. direct evidence. 
The reputation vector ${\bf r}\in [0, 1]^n$ is defined as a function of $\alpha$, ${\bf s}$ and $A$ by the following equation:
\begin{equation}
	r_x = (1 - \alpha) s_x + \alpha \sum_{y=1}^n \frac{r_y}{\ell}A_{yx}
\label{reputationrA}
\end{equation} 
where $\ell = \sum_{z=1}^n r_z $.
\end{example}

Eq.~(\ref{reputationrA}) can be read as follows. 
To determine the reputation of user $x$ we first take into account the direct information about $x$. 
From this we can compute $s_x$, the reputation initially assigned to $x$ if no further information is available. 
However, additional information may be available, namely the aggregated ratings in $A$. 
The weight of direct versus indirect information 
is accounted for by the parameter $\alpha$. 
If no direct information about $x$ is available, the reputation of $x$ can be computed as $r_x = \sum_y(r_y / \ell)A_{yx}$, i.e.\ a weighted average of the reputation values $A_{yx}$ with weights equal to the normalized reputations. 
Adding the two contributions, with weights $\alpha$ and $1-\alpha$, 
yields (\ref{reputationrA}): a weighted average over all available information. 

The equation for solving the unknown ${\bf r}$ contains $\bf r$.
A solution is obtained by repeatedly substituting (\ref{reputationrA}) into itself
until a stable state has been reached. This is the steady state of the Markov chain.
It was shown in \cite{flowbase} that Eq.~(\ref{reputationrA}) always has a solution and that the solution is unique.

Intuitively, $A$ can be seen as an adjacency matrix of a trust network where nodes represent entities and edges represent the direct trust that entities have in other entities based on direct experience.
Based on the results presented in \cite{flowbase}, Eq.~(\ref{reputationrA}) can be applied to assess reputation for arbitrary trust networks.


\subsection{Subjective Logic}
\label{sec:prelimsubjective}

Subjective Logic (SL)  is a trust algebra based on Bayesian theory 
and Boolean logic that explicitly models uncertainty and belief ownership.
In the remainder of this section, we provide an overview of SL based on \cite{josang2001logic}.

The central concept in SL is the three-component {\em opinion}.

\begin{definition}[Opinion and opinion space] \cite{josang2001logic}
\label{def:opinion}
An opinion $x$ about some proposition $P$
is a tuple $x=(\xb,\xd,\xu)$,
where $\xb$ represents the belief that $P$ is provable (\emph{belief}),
$\xd$ the belief that $P$ is disprovable (\emph{disbelief}),
and $\xu$ the belief that $P$ is neither provable nor disprovable (\emph{uncertainty}).
The components of $x$ satisfy $\xb+\xd+\xu=1$.
The space of opinions is denoted as $\Omega$ and is defined as
$\Omega=\{(b,d,u)\in[0,1]^3 \; |\;  b+d+u=1\}$.
\end{definition}

An opinion $x$ with $\xb+\xd<1$
can be seen as an incomplete probability distribution.
In order to enable the computation of expectation values,
SL extends the three-component opinion with a fourth parameter `$a$'
called relative atomicity, with $a\in[0,1]$.
The probability expectation is $E(x)=\xb+\xu a$.
In this paper, we will omit the relative atomicity from our notation,
because in our context (trust networks) it is not modified by
any of the computations on opinions.
In more complicated situations, however, the relative atomicity
is modified in nontrivial ways.


Opinions are based on evidence.
Evidence can be represented as a pair of nonnegative finite numbers $(p,n)$,
where $p$ is the amount of evidence supporting the proposition,
and $n$ the amount that contradicts  the proposition.
The notation $e=p+n$ is used to denote the total amount of evidence about the proposition.
There is a one-to-one mapping between an opinion $x\in\Omega$ and the evidence $(p,n)$
on which it is based,
\begin{equation}
\label{mappingJosang}
	(\xb,\xd,\xu)=\frac{(p,n,2)}{p+n+2};
	\quad
	(p,n)=\frac{2(\xb,\xd)}{\xu}.
\end{equation}
The bijection~(\ref{mappingJosang}) holds for any value of the atomicity~$a$.
It has its origin in an analysis of the a posteriori probability distribution
(Beta function distribution) 
of the biases which underlie the generation of evidence \cite{josang2001logic}.
This distribution is given by
\begin{equation}
	\rho(t|p,n,a)=\frac{t^{-1+p+2a}(1-t)^{-1+n+2(1-a)}}{B(p+2a,n+2-2a)},
\label{betadist}
\end{equation}
where $t$ is the probability that proposition $P$ is true, and
$B(\cdot,\cdot)$ is the Beta function.
The opinion $x=\frac{(p,n,2)}{p+n+2}$ is based on the vague knowledge (\ref{betadist}) about~$t$.
The left part of (\ref{mappingJosang}) holds because the thus constructed opinion $x$ 
has the same expectation as (\ref{betadist}).

Intuitively, the amount of positive and negative evidence about a proposition determines the belief and the disbelief in the proposition, respectively.
Increasing the total amount of evidence ($e$) reduces the uncertainty.
Note that there is a fundamental difference between an opinion where a proposition is equally provable and disprovable and one where we have complete uncertainty about the proposition.
For instance, opinion $(0,0,1)$ indicates that there is no evidence either supporting or contradicting the proposition, i.e. $n=p=0$, 
whereas opinion $(0.5,0.5,0)$ indicates that $n=p=\infty$.

%
We use the notation $p(x)\stackrel{\rm def}{=}2\frac{\xb}{\xu}$ to denote the
amount of supporting evidence underlying opinion~$x$,
and likewise $n(x)\stackrel{\rm def}{=}2\frac{\xd}{\xu}$ for the amount of `negative' evidence.
Moreover, we use the notation $e(x)=p(x)+n(x)$ to represent the total amount of evidence underlying opinion~$x$.




SL provides a number of operators to combine opinions.
One of the fundamental operations is the combination
of evidence from multiple sources.
Consider the following scenario.
Alice has evidence $(p_1,n_1)$ about some proposition. She forms an opinion
$x_1=(p_1,n_1,2)/(p_1+n_1+2)$.
Later she collects additional evidence $(p_2,n_2)$ independent of the first evidence.
Based on the second body of evidence alone, she would arrive at opinion
$x_2=(p_2,n_2,2)/(p_2+n_2+2)$. If she combines all the evidence, she obtains opinion
\begin{equation}
	x=\frac{(p_1+p_2,n_1+n_2,2)}{p_1+p_2+n_1+n_2+2}.
\label{propaddedev}
\end{equation}
The combined opinion $x$ is expressed as a function of $x_1$ and $x_2$ via the so-called
`consensus' rule; this is denoted as $x=x_1\oplus x_2$.

\begin{definition}[Consensus] {\bf \cite{josang2001logic}}
\label{def:consensus}
Let $x,y\in\Omega$. 
The consensus $x\oplus y\in\Omega$ is defined as
\begin{equation}
	x\oplus y \stackrel{\rm def}{=}
	\frac{(\xu \yb+\yu \xb,\xu \yd+\yu \xd,\xu\yu)}{\xu+\yu-\xu\yu}.
\label{eqconsensus}
\end{equation}
\end{definition}
Eq.~(\ref{eqconsensus}) precisely corresponds to (\ref{propaddedev}).
Note that $x\oplus y=y\oplus x$.
Furthermore, the consensus operation is associative, i.e.
$x\oplus(y\oplus z)=(x\oplus y)\oplus z$.
These properties are exactly what one intuitively expects from an operation
that combines evidence.
It is worth noting that the evidence has to be {\em independent} for the $\oplus$ rule to apply.
Combining dependent evidence would lead to the problem of double counting evidence.
We formalize and discuss this problem in Section~\ref{sec:issue-applicationTN}.


The second important operation in SL is the transfer of opinions from one party to another.
Consider the following scenario.
Alice has opinion $x$ about Bob's trustworthiness.
Bob has opinion $y$ about some proposition~$P$.
He informs Alice of his opinion. Alice now has to form an opinion about~$P$.
The standard solution to this problem is that Alice applies an $x$-dependent weight
to Bob's opinion~$y$ \cite{Bhuiyan10,kamvar2003eigentrust,Ching10,Richters11,flowbase}.
This is the so called `discounting'.
The following formula is usually applied.

\begin{definition}[Discounting] {\bf \cite{josang2001logic}}
\label{def:discounting}
Let $x,y\in\Omega$.
The discounting of opinion $y$ using opinion $x$ is denoted as $x\otimes y\in\Omega$,
and is defined as
\begin{equation}
	x\otimes y \stackrel{\rm def}{=}(\xb\yb,\; \xb\yd,\; \xd+\xu+\xb\yu).
\label{otimes}
\end{equation}
\end{definition}
It holds that $x\otimes y\neq y\otimes x$ and that $x\otimes(y\otimes z)=(x\otimes y)\otimes z$.
The discounting rule (\ref{otimes}) is {\em not} distributive w.r.t. consensus, i.e.
$(x\oplus y)\otimes z\neq (x\otimes z)\oplus(y\otimes z)$
and $x\otimes(y\oplus z)\neq (x\otimes y)\oplus(x\otimes z)$.


In SL, a trust network
can be modeled with a combination of consensus and discounting operators.
The consensus operator is used to aggregate trust information from different sources, while the discounting operator is used to implement  trust transitivity.
Note that, in a trust network, SL distinguishes two types of trust relationship: \emph{functional trust}, which represents the opinion about an entity's ability to provide a specific function, and \emph{referral trust}, which represents the opinion about an entity's ability to provide 
recommendations about other entities.
Referral trust is assumed to be transitive, and a trust chain is said to be valid if the last edge of the chain represents functional trust and all previous edges represent referral trust.




\section{Limitations of Subjective Logic}
\label{sec:flowSL}

Our desideratum is a novel reputation metric that has the advantages of both SL and flow-based models.
On one hand, we aim at an automated procedure for computing reputation as in flow-based approaches.
On the other hand, we aim to determine the confidence in reputation values by making uncertainty explicit as in SL.
In this section, we discuss the limitations of SL.
Then, in Section~\ref{sec:combining-FB-SL} we show how these limitations affect a na{\"i}ve approach that combines flow-based reputation and SL.

\subsection{Dogmatic Opinions}
\label{sec:problemdogmatic}

\begin{definition}
\label{def:BDU}
The extreme points corresponding to full belief ($B$), full disbelief ($D$) 
and full uncertainty ($U$)
are defined as
\begin{eqnarray}
	B\stackrel{\rm def}{=}(1,0,0) & \quad
	D\stackrel{\rm def}{=}(0,1,0) & \quad
	U \stackrel{\rm def}{=}(0,0,1).
\end{eqnarray}
\end{definition}
The special points $B,D,U$ behave as follows regarding the consensus operation:
$B\oplus x=B$; $D\oplus x=D$; $U\oplus x=x$.
The full uncertainty $U$ behaves like an additive zero.

With respect to the discounting rule,
the special points $B,D,U$ behave as $B\otimes x=x$, $D\otimes x=U$, $U\otimes x=U$,
$x\otimes U=U$, $x\otimes B=(\xb,0,1-\xb)$,
$x\otimes D=(0,\xb,1-\xb)$.

Opinions that have $u=0$ (i.e., lying on the line between $B$ and $D$) are called `dogmatic'
opinions.
They have to be treated with caution,
since they have $e=\infty$ and therefore overwhelm other opinions when the consensus $\oplus$ is applied.
We will come back to this issue in Section~\ref{sec:excludedogmatic}.

\subsection{Counter-intuitive behaviour of the $\otimes$ operation}
\label{sec:flaw}

We observe that the discounting rule $\otimes$ does not have a natural
interpretation in terms of evidence handling.
For instance, if we compute the positive evidence contained in $x\otimes y$
we get, using $p,n$ notation as introduced in Section~\ref{sec:prelimsubjective},
\begin{eqnarray}
	p(x\otimes y)&=& 2\frac{(x\otimes y)_{\rm b}}{(x\otimes y)_{\rm u}}=
	2\frac{\xb\yb}{\xd+\xu+\xb\yu} =
	2\frac{\frac14 p(x)p(y)}{\frac12 n(x)/\yu+1/\yu+\frac12 p(x)}
	\nonumber\\ &=&
	\frac{p(x)p(y)}{[2+n(x)][1+2p(y)+2n(y)]+p(x)}
\label{pxy}
\end{eqnarray}
where in the final step we have used $1/\yu=1+2p(y)+2n(y)$.
Similarly,  we get
\begin{equation}
	n(x\otimes y)=2\frac{(x\otimes y)_{\rm d}}{(x\otimes y)_{\rm u}}=
	\frac{p(x)n(y)}{[2+n(x)][1+2p(y)+2n(y)]+p(x)}.
\label{nxy}
\end{equation}
Eqs.~(\ref{pxy}) and~(\ref{nxy}) are complicated functions of the amounts $p(x), n(x), p(y), n(y)$.
We show that 
they do not have a clear and well-defined interpretation in terms of evidence handling.
For instance, Eq.~(\ref{pxy}) does not allow us to determine whether the evidence underlying  $x\otimes y$ originates from $x$ or $y$;
one can argue that $p(x\otimes y)$ is either a ``fraction'' of $p(y)$, a ``fraction'' of $p(x)$, or contains evidence from both $y$ and $x$.
Even if we try to interpret in the standard way (see Section~\ref{sec:prelimsubjective}), we can observe that the factor multiplying $p(y)$ in (\ref{pxy}) and $n(y)$ in (\ref{nxy}) depends on $p(y)$ and $n(y)$.
Hence, the evidence underlying $x\otimes y$ is not simply an $x$-dependent multiple of
the evidence underlying~$y$.
In fact, the examples below show that in some cases the contribution from $y$ 
completely disappears from the equation.

\begin{example}
\label{ex:strangelimit}
$\phantom{.}$ \newline
Let $x,y\in\Omega$, with $n(x)=0$, $n(y)=0$ and $p(y)\gg p(x)$.
Then $p(x\otimes y)\approx p(x)/4$.
\end{example}

\begin{example}
\label{ex:strangelimit2}
$\phantom{.}$ \newline
Let $x,y\in\Omega$, with $n(x)=0$, $p(y)=0$ and $n(y)\gg p(x)$.
Then $n(x\otimes y)\approx p(x)/4$.
\end{example}
In both these examples, $y$ is based on a lot of evidence;
but even if $x$ contains a lot of belief, none of $y$'s evidence survives in $x\otimes y$.
We conclude that the discounting operation~$\otimes$
gives counter-intuitive results.


The $\otimes$ rule is inspired by a {\em probabilistic} interpretation
of opinions.
%
The probabilistic interpretation might suggest that it is natural to multiply probabilities,
i.e. that the expressions
$(x\otimes y)_{\rm b}=\xb\yb$ and $(x\otimes y)_{\rm d}=\xb\yd$
are intuitively correct.
However, we argue that this is not at all self-evident.
When discounting $y$ through $x$, the uncertainties in $x$
induce an $x$-dependent probability distribution on~$y$.
This can be thought of as an additional layer of uncertainty about
beta distributions. 
Let (\ref{betadist}) describe  opinion~$y$;
then the discounting through $x$ introduces uncertainty about
the parameters $p$ and $n$ in the equation (a probability distribution on $p$ and $n$).
It is not at all self-evident that the resulting opinion is $x\otimes y$
as prescribed by Def.~\ref{def:discounting}.
It would make equal sense to replace
the discounting factor $x_{\rm b}$ by e.g. the expectation $x_{\rm b}+ax_{\rm u}$. 
In this paper we do not pursue such an approach based on distributions, 
but we mention it in order to point out
that the $\otimes$ rule is not necessarily well-founded.





The fact that SL employs on the one hand
a consensus rule based on {\em adding evidence}
and on the other hand
a discounting rule based on {\em multiplying opinions}
leads to a lack of `cooperation' between the
$\oplus$ and $\otimes$ operations.
Most importantly, the $\otimes$ is not distributive with respect to $\oplus$,
\begin{equation}
	(x\otimes y)\oplus (x\otimes z) \neq x\otimes(y\oplus z).
\label{nodistribution}
\end{equation}
Consider the following scenario. 
%
\begin{example}
\label{ex:nodistrib}
Alice has opinion $x$ about Bob's trustworthiness in providing recommendations.
Bob experiments with chocolate on Monday and forms an opinion $y$ about its medicinal qualities.
On Tuesday he does some more of the same kind of experiments and forms an independent opinion $z$. 
He informs Alice of $y$, $z$ and his final opinion $y\oplus z$.
What should Alice think about the medicinal qualities of chocolate?
One approach is to say that Alice should appraise opinions $y$ and $z$ separately,
yielding $(x\otimes y)\oplus (x\otimes z)$ (Note that the two occurrences of $x$ represent the very same opinion, i.e.\ the evidence underlying the two occurrences is the same).
Another approach is to weight Bob's combined opinion, yielding $x\otimes(y\oplus z)$.
Intuitively the two approaches should yield exactly the same opinion,
yet the SL rules give (\ref{nodistribution}).
%
\end{example}


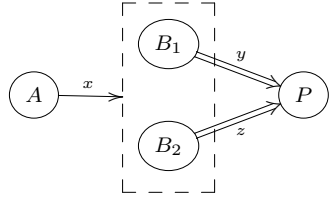
\begin{figure}[!t]
\centering
{\small
$
\def\g#1{\save
[].[dddd]!C="g#1"*[F--]\frm{}\restore}%
\xymatrix @R=0.1pc {
& & \g1 \hspace{1cm} &\\
& & *++[o][F-]{B_1} \ar@{=>}[rd]^y \\
& *++[o][F-]{A}  &  & *++[o][F-]{P} \\
& &*++[o][F-]{B_2}\ar@{=>}[ru]_z\\
& & 
\ar@{->}^(.4){x} "3,2" ; "g1"
}
$
}
\caption{\it
Trust network representing the scenario in Example~\ref{ex:nodistrib}.
$A$ is Alice, $B_1$ and $B_2$ are Bob on Monday and Tuesday respectively, and $P$ is the proposition ``chocolate has medicinal qualities''.
Referral trust is drawn as a full line, and functional trust as a double full line.}
\label{fig:network1}
\end{figure}

We now present a numerical example that illustrates the issue discussed above.
\begin{example}\label{ex:nodistrib2}
Figure~\ref{fig:network1} shows the trust network representing the scenario in Example~\ref{ex:nodistrib}.
To highlight that opinions $x$ and $y$ are independent, in the figure we abuse the network notation and duplicate the node representing Bob: $B_1$ represents Bob on Monday and $B_2$ represents Bob on Tuesday.
The edge between Alice ($A$) and Bob (dashed rectangle) represents Alice's opinion $x$ about Bob's recommendations.
This opinion concerns Bob's recommendations regardless of when they are formed (e.g., on Monday or on Tuesday). 

Suppose that Alice's opinion about Bob's trustworthiness is $x=(0.6,0.1,0.3)$, and Bob's opinions about the proposition $P$ 
are $y=(0.3,0.6,0.1)$ and $z=(0.5,0.2,0.3)$.
We are interested in Alice's opinion $w$ about $P$ based on Bob's recommendations.
As discussed in the previous example, we have two approaches to compute such an opinion:
\begin{enumerate}
\item $w=(x\otimes y)\oplus (x\otimes z)=(0.314,0.341,0.345)$
\item $w=x\otimes(y\oplus z)=(0.227,0.324,0.449)$
\end{enumerate}
Clearly, the two approaches yield different opinions, contradicting the intuitive expectation. 
\end{example}

\subsection{Double counting of evidence in Trust Networks}
\label{sec:issue-applicationTN}

%
%
%

The $\oplus$ rule imposes constraints on the evidence that can be aggregated: 
it requires evidence to be independent \cite{josang2001logic}.
In the literature, however, there is no well-defined notion of evidence independence.
Some researchers \cite{Sentz02,Zomlot:2011} assume that pieces of evidence are independent if they are obtained from independent sources, 
where two sources are said to be independent if they measure completely unrelated features.
This definition, however, is too restrictive.
For instance, the evidence collected by a sensor 
at different points 
in time can also be independent.

Evidence is usually extracted from ``observations'' of a system.
In this work we adopt a notion of evidence independence based on the independence of observations,
which we define in the same way as independence of random variables.

\begin{definition}[Independent Evidence]
Let $o_i$ and $o_j$ be two observations and $e_i$ and $e_j$ the evidence obtained from $o_i$ and $o_j$ respectively.
We say that $o_i$ and $o_j$ are independent if and only if ${\rm Prob}(o_i|o_j) = {\rm Prob}(o_i)$.
Moreover, we say that $e_i$ and $e_j$ are independent if they are obtained from independent observations.
\end{definition}
Intuitively, observation independence requires that the probability that $o_i$ happens, assuming that  $o_j$ happened, is the same as the probability that $o_i$ happens regardless of $o_j$.
The definition above can be extended to opinions:
\begin{definition}[Independent Opinions]
Let $x,y\in\Omega$ be opinions.
We say that $x$ and $y$ are independent if and only if the evidence underlying $x$
and $y$ is independent.
\end{definition}

Combining dependent evidence leads to the problem of double counting evidence.
\begin{definition}[Double Counting]
\label{def:doublecount}
Let $x,y\in\Omega$.
In an expression of the form $x \oplus y$
we say that there is double counting if 
there is dependence between the evidence underlying $x$ and the evidence underlying~$y$.
\end{definition}
Intuitively, dependent evidence shares ``part'' of the evidence.
Therefore, aggregating dependent evidence leads to counting some part of evidence more than once.


\begin{example}
\label{ex:simple}
Consider the expression $(y\otimes x)\oplus(z\otimes x)$, where both occurrences of $x$ are obtained from the same observation.
The evidence underlying $x$ is contained on the left side as well as the right side of the `$\oplus$'.
This is a clear case of double counting.
\end{example}

\begin{example}
\label{ex:oplustransport}
Consider the expression $(x\otimes y)\oplus(x\otimes z)$, again with both instances `$x$' coming from the
same observation.
The evidence underlying $x$ is contained on the left side as well as the right side of the `$\oplus$',
but less evidently than in Example~\ref{ex:simple}, because now $x$ is used for discounting.
In Section~\ref{sec:flaw} we showed that the $\otimes$ rule causes evidence from $x$
to end up in $x\otimes y$ in a complicated way. Hence the opinions $x\otimes y$ and
$x\otimes z$ are {\em not} independent, which causes double counting of $x$ in the expression
$(x\otimes y)\oplus(x\otimes z)$.
\end{example}
It is worth noting that double counting of $x$ in the expression $(x\otimes y)\oplus(x\otimes z)$ can also be observed in Example~\ref{ex:nodistrib2}.
Indeed, the uncertainty in $(x\otimes y)\oplus(x\otimes z)$ is lower than the uncertainty in $x \otimes (y\oplus z)$, 
indicating that the result contains more evidence when the trust network is represented using the first expression.

To avoid the problem of double counting, SL requires that the trust network is expressed in a \emph{canonical form} \cite{Josang2006simplification,Josang:2006}, where all trust paths are independent.
Intuitively, a trust network expression is in canonical form if every edge appears only once in the expression.
\begin{example}
Consider the two trust network expressions representing the trust network in Figure~\ref{fig:network1} given in Example~\ref{ex:nodistrib}: $(x\otimes y)\oplus (x\otimes z)$ and $x\otimes(y\oplus z)$.
The first expression is not in canonical form as opinion $x$ occurs twice in the expression; the second expression is in canonical form as every edge appears only once in the expression.
Thus, $x\otimes(y\oplus z)$ is the proper representation of the trust network in Figure~\ref{fig:network1}.
\end{example}
In the next section we show that it is not always possible to express a trust network in canonical form.
As suggested in \cite{Josang2006simplification,Josang:2006}, this issue can be addressed by removing some edges from the network.
This, however, means discarding part of the trust information, thus reducing the quality of 
reputation values.

\section{Combining flow-based reputation and Subjective Logic}\label{sec:combining-FB-SL}

This section presents a na{\"i}ve approach that combines flow-based reputation 
and SL. We illustrate the limitations of such a na{\"i}ve approach.

We first introduce some notation and definitions.
Flow-based reputation models usually assume that users  who are honest during transactions are also honest in reporting their ratings \cite{kamvar2003eigentrust}.
This assumption, however, does not hold in many real-life situations \cite{Abdul-Rahman:1998}.
Thus, as it is done in SL, we distinguish between referral trust and functional trust (see Section~\ref{sec:preliminaries}).
We use $A$ to represent direct referral trust and $T$ to represent direct functional trust.
$R$ denotes the final referral trust and $F$ the final functional trust.

\begin{definition}
For $n$ users, the \emph{direct referral trust matrix} $A$ is an $n\times n$ matrix, where
$A_{xy} \in \Omega$ (with $x\neq y$) is the direct referral trust that user $x$ has in user $y$, and
$A_{xx} = (0,0,1)$ for all~$x$.
\end{definition}
Note that we impose the condition $A_{xx} = (0,0,1)$ in order to prevent artifacts
caused by self-rating \cite{flowbase}.

Let $T_{jP}$ be the opinion of user $j$ about some proposition $P$, 
and let $R_{ij}$ be $i$'s (possibly indirect) opinion about the trustworthiness of user $j$.
The opinion of user $i$ about $P$ based on direct and indirect evidence can be computed using the following equation:
\begin{equation}
\label{eq:functional+sub}
	F_{iP} = T_{iP} \oplus \bigoplus_{j: j \neq i} (R_{ij} \otimes T_{jP}).
\end{equation}
Eq.~(\ref{eq:functional+sub}) computes the final functional trust $F_{iP}$ by combining 
user $i$'s direct opinion $T_{ip}$ with all the available opinions of other users, $\{T_{jP}\}_{j\neq i}$.
The opinion received from~$j$ is weighted with the `reputation' $R_{ij}$.

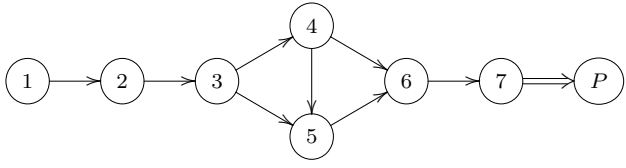
\begin{figure}[!t]
\centering
{\small
$
\xymatrix @R=0.3pc @C=1.6pc {
            && & *++[o][F-]{4}\ar[dd]\ar[dr] &  \\
 *++[o][F-]{1} \ar[r] &  *++[o][F-]{2} \ar[r] & *++[o][F-]{3} \ar[ru]\ar[rd] &          & *++[o][F-]{6} \ar[r]& *++[o][F-]{7}\ar@{=>}[r]& *++[o][F-]{P}\\
             &&    & *++[o][F-]{5}\ar[ur] &  
}
$
}
\caption{\it Example of a trust network that is problematic for Subjective Logic.} 
\label{fig:problem1}
\end{figure}

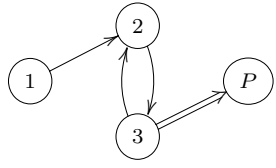
\begin{figure}[!t]
\centering
{\small
$
\xymatrix @R=0.3pc {
                 & *++[o][F-]{2}\ar@/^/ @{-{>}} [dd] &  \\
*++[o][F-]{1} \ar[ru] &          &  *++[o][F-]{P}\\
                 & *++[o][F-]{3}\ar@/^/ @{-{>}}[uu] \ar@{=>}[ru]&  \\
}
$
}
\caption{\it Example of a trust network containing a loop.}
\label{fig:loop123}
\end{figure}

To find $R_{ij}$, we could try a recursive approach\footnote{
We consider only scenarios where all entities {\em publish} their opinions.
If opinions are not published but communicated solely over the network links,
then a recursive equation containing $A\otimes R$ (the opposite order of $R\otimes A$)
applies.} 
inspired by Eq.~(\ref{reputationrA}):
\begin{equation}
\label{eq:flow+sub}
	R_{ij} = A_{ij} \oplus \bigoplus_{k: k \neq i} (R_{ik} \otimes A_{kj})
	\quad\quad \mbox{ for }i\neq j.
\end{equation}
To demonstrate the problems that occur in this na\"ive
combination of flow-based reputation and SL, we consider 
the trust networks shown in Figures~\ref{fig:problem1} and~\ref{fig:loop123}.
Referral trust is drawn as a full line, functional trust as a double full line.

Figure~\ref{fig:problem1} is a variant of a network discussed in \cite{Josang2006simplification}.
We are interested in determining the opinion $F_{1P}$ of user~$1$ about some proposition $P$.
User~$1$ does not have any direct evidence.
The only functional trust about $P$ comes from user~$7$.
Thus, we have to determine $R_{17}$, user $1$'s referral trust in user~$7$.
This is done by taking $A_{67}$ with the proper weight, namely $R_{16}$.
Continuing this recursive approach using (\ref{eq:flow+sub}) gives
\begin{equation}
\begin{array}{l@{\ }l}
F_{1P} = &[A_{12} \otimes A_{23} \otimes A_{34} \otimes A_{46} \oplus (A_{12} \otimes A_{23} \otimes A_{34} \otimes A_{45}\\
& \oplus A_{12} \otimes A_{23} \otimes A_{35}) \otimes A_{56}] 
\otimes A_{67} \otimes T_{7P}.
\end{array}
\label{F1P}
\end{equation}
This, however, is a problematic result.
Recall that SL requires trust networks to be expressed in a canonical form. 
If this restriction is not satisfied, we face the problem of `double counting' opinions,
i.e. applying the $\oplus$ operation to opinions that are not independent (Def.~\ref{def:doublecount}).

In Figure~\ref{fig:problem1},
consider the case $A_{45}=U$.
The canonical solution for this case is 
\begin{equation}
\label{eq:canonical}
F_{1P}=A_{12}\otimes A_{23}\otimes(A_{34}\otimes A_{46}\oplus A_{35}\otimes A_{56})
\otimes A_{67}\otimes T_{7P}
\end{equation}
whereas (\ref{F1P}) yields
\begin{equation}
\label{eq:flow-SL-U}
F_{1P}=(A_{12}\otimes A_{23}\otimes A_{34}\otimes A_{46}
\oplus A_{12}\otimes A_{23}\otimes A_{35}\otimes A_{56})\otimes A_{67}\otimes T_{7P}.
\end{equation}
In (\ref{eq:flow-SL-U})
the $A_{12}$ and $A_{23}$ are double-counted.
We conclude that the na\"ive recursive equation (\ref{eq:flow+sub}) does not properly reproduce
the canonical solution.

There is a further problem, unrelated to the na\"ive recursive approach.
As was shown in \cite{Josang2006simplification}, it is not even possible to transform the trust network in
Figure~\ref{fig:problem1} into a canonical form in the general case $A_{45}\neq U$.

The problems become even worse when the trust network contains loops,
e.g. a loop as shown in Figure~\ref{fig:loop123}.
Here too, there is no canonical form.
Applying the recursive approach to Figure~\ref{fig:loop123} gives
$R_{13}=R_{12}\otimes A_{23}$, with
$R_{12}=A_{12}\oplus R_{12}\otimes A_{23}\otimes A_{32}$.
Repeatedly substituting the latter into itself yields
\begin{equation}
	R_{12}=A_{12}\oplus [A_{12}\oplus \{A_{12}\oplus\cdots\}\otimes A_{23}\otimes A_{32}]
	\otimes A_{23}\otimes A_{32}.
\label{loopSL}
\end{equation}
We observe that taking opinion $A_{32}$ into account causes excessive double-counting of~$A_{12}$.
If the loop is somehow discarded, then the information contained in $A_{32}$ is destroyed.

In conclusion,
(i)
generic trust networks with several connections $A_{ij}\neq U$
cannot be handled with SL because there is no canonical form for them
that avoids double-counting;
(ii)
even when there is a canonical result, this result cannot be reproduced 
by a straightforward recursive approach.

\section{Subjective Logic revisited}
\label{sec:SLrevisited}


This section presents a new, fully evidence-based approach to SL.
We refer to the resulting opinion algebra as Evidence-Based Subjective Logic
or EBSL.

\subsection{Excluding dogmatic opinions}
\label{sec:excludedogmatic}

As mentioned in Section~\ref{sec:problemdogmatic},
dogmatic opinions are problematic when the $\oplus$ operation is applied
to them.
Furthermore, a dogmatic opinion corresponds to an infinite amount of evidence,
which in our context is not realistic.
In the remainder of this paper, we will exclude dogmatic opinions.
We will work with a reduced opinion space defined as follows.

\begin{definition}
The \emph{opinion space excluding dogmatic opinions} is denoted as $\Omega'$ and is defined as
$\Omega'\stackrel{\rm def}{=}\{(b,d,u)\in[0,1)\times[0,1)\times(0,1] \; |\;  b+d+u=1\}$.
\end{definition}
We are by no means the first to do this; in fact, the exclusion of
dogmatic opinions was proposed as an option in the very early literature on SL \cite{Josang:2006}.

\subsection{The relation between evidence and opinions: a simplified justification}
\label{sec:evidencemap}

%

We make a short observation about the mapping
between evidence and opinions.
As was mentioned in Section~\ref{sec:prelimsubjective}, 
there is a one-to-one mapping (\ref{mappingJosang})
based on the analysis of probability distributions (Beta distributions).
Here we show that there is a shortcut:
the same mapping can also be obtained in a much simpler way,
based on constraints. 

\begin{theorem}
\label{th:simplemap}
Let $p\geq0$ be the amount of evidence that supports `belief';
let $n\geq0$ be the amount of evidence that supports `disbelief'.
Let $x=(b,d,u)\in\Omega'$ be the opinion based on the evidence.
If we demand the following four properties,
\begin{enumerate}
\item
$b/d=p/n$
\item
$b+d+u=1$
\item
$p+n=0 \;\Rightarrow\; u=1$
\item
$p+n\to\infty \;\Rightarrow\; u\to 0$
\end{enumerate}
then the relation between $x$ and $(p,n)$ has to be
\begin{eqnarray}
	x=(b,d,u)=\frac{(p,n,c)}{p+n+c} &;&\quad
	(p,n)=c\frac{(b,d)}{u}
\label{evidence}
\end{eqnarray}
where $c>0$ is a constant.
\end{theorem}
\underline{\it Proof:}
Property~1 gives $(b,d)=(p,n)/f(p,n)$ where $f$ is some function.
Combined with property~2 we then have $\frac{p+n}{f(p,n)}=1-u$.
Property~3 then gives $\frac{0}{f(0,0)}=0$, while
property~4 gives $\lim_{p+n\to\infty} \frac{p+n}{f(p,n)}=1$.
The latter yields $f(p,n)=p+n+c$ where $c$ is some constant.
Allowing $c<0$ would open the possibility of components of $x$ being negative.
Thus we must have $c\geq 0$.
The requirement $\frac{0}{f(0,0)}=0$ eliminates the possibility of having
$c=0$, since $c=0$
would yield division by zero.
Finally, setting $u=c/(p+n+c)$ is necessary to satisfy property~2.
\hfill$\square$

Theorem~\ref{th:simplemap} shows that we can derive a formula similar to
(\ref{mappingJosang}), based on minimal requirements which make intuitive sense.
Only the constant $c$ is not fixed by the imposed constraints;
it has to be determined from the context.
One can interpret $c$ as a kind of soft threshold on the amount of evidence:
beyond this threshold one starts gaining enough confidence from the evidence to form
an opinion.

We observe that (\ref{evidence}) with its generic constant~$c$ is already sufficient to
derive the consensus rule $\oplus$, i.e. {\em the consensus rule does not require $c=2$}.

\begin{lemma}
\label{lemma:consensus}
The mapping (\ref{evidence}) with arbitrary $c$
implies the consensus rule $\oplus$ as specified in
Def.~\ref{def:consensus}.
\end{lemma}
\underline{\it Proof:}
Consider $x=(b_1,d_1,u_1)=(p_1,n_1,c)/(p_1+n_1+c)$
and $y=(b_2,d_2,u_2)=(p_2,n_2,c)/(p_2+n_2+c)$.
An opinion formed from the combined evidence $(p_1+p_2,n_1+n_2)$
according to (\ref{evidence}) is given by
$(b,d,u)=\frac{(p_1+p_2,n_1+n_2,c)}{p_1+n_1+p_2+n_2+c}$.
Substituting $p_i=c \frac{b_i}{u_i}$ and $n_i=c \frac{d_i}{u_i}$ yields, after some simplification,
$(b,d,u)=\frac{(u_1 b_2+u_2 b_1,u_1 d_2+u_2 d_1,u_1 u_2)}{u_1+u_2-u_1 u_2}$.
\hfill$\square$
Theorem~\ref{th:simplemap} and Lemma~\ref{lemma:consensus}
improve our understanding of evidence-based opinion forming
and of the consensus rule.

Furthermore, in (\ref{mappingJosang}) and (\ref{betadist})
we can replace `$2$' by $c$ and the expectation of $t$,
obtained by integrating the Beta distribution times $t$, is still 
$x_{\rm b}+ax_{\rm u}$!
Therefore, in the remainder of the paper, we will work with a re-defined version
of the $p(x)$ and $n(x)$ functions (Section~\ref{sec:prelimsubjective}).
The new version has a general value $c>0$ instead of $c=2$.

\begin{definition}
\label{def:ev}
Let $x=(\xb,\xd,\xu)\in\Omega'$. We define the notation $p(x)$, $n(x)$
and $e(x)$ as
\begin{equation}
	p(x)\stackrel{\rm def}{=} c\frac{\xb}{\xu}
	; \quad	
	n(x)\stackrel{\rm def}{=} c\frac{\xd}{\xu}
	; \quad
	e(x)\stackrel{\rm def}{=} p(x)+n(x).
\label{defev}
\end{equation}
\end{definition}


\subsection{Scalar multiplication}
\label{sec:scalar}

Our next contribution has more impact.
We define an operation on opinions that is equivalent
to a scalar multiplication on the total amount of evidence.

\begin{definition}[Scalar multiplication]
\label{def:mult}
Let $x=(\xb,\xd,\xu)\in\Omega'$ and let $\alpha\geq 0$ be a scalar.
We define the product $\alpha\cdot x$ as
\begin{equation}
	\alpha \cdot x \stackrel{\rm def}{=} \frac{(\alpha \xb,\alpha \xd,\xu)}{\alpha(\xb+\xd)+\xu}.
\label{defmult}
\end{equation}
\end{definition}



\begin{lemma}
\label{lemma:scalarprop}
Let $x\in\Omega'$ and $\alpha\geq 0$.
The scalar multiplication as specified in Def.~\ref{def:mult}
has the following properties:
\begin{enumerate}
\item
$\alpha\cdot x\in\Omega'$.
\item
$0\cdot x=U$.
\item
$1\cdot x=x$.
\item
For $n\in\mathbb N$, $n\geq 2$, it holds that $n\cdot x=\underbrace{x\oplus x\oplus\cdots\oplus x}_{n}$.
\item
The evidence underlying $\alpha\cdot x$
is $\alpha$ times the evidence underlying~$x$, i.e.
$p(\alpha\cdot x)=\alpha p(x)$ and $n(\alpha\cdot x)=\alpha n(x)$.
\item
If $\alpha\neq 0$ then
$(\alpha\cdot x)_{\rm b}/(\alpha\cdot x)_{\rm d}=\xb/\xd$.
\end{enumerate}
\end{lemma}
\underline{\it Proof:}
Property~1:
It is readily verified that the components add to~1, using $\xb+\xd=1-\xu$.
Since $\alpha\geq 0$, all three components of $\alpha\cdot x$ are nonnegative, since
$x\in\Omega'$.
Properties 2 and~3: Found directly by substituting $\alpha=0$ resp. $\alpha=1$ in (\ref{defmult}).
Property~4: Consider $n=2$. Setting $\alpha=2$ in (\ref{defmult}) yields
$2\cdot x=\frac{(2\xb,2\xd,\xu)}{2-\xu}=\frac{(2\xu\xb,2\xu\xd,\xu^2)}{2\xu-\xu^2}$
$=x\oplus x$. The rest follows by induction.
Property~5: We use (\ref{evidence}) to map between opinions and evidence.
The positive evidence of $x$ is $p(x)=c\xb/\xu$.
The positive evidence of $\alpha\cdot x$ is $p(\alpha\cdot x)$
$=c\frac{\alpha\xb}{\alpha(1-\xu)+\xu}/\frac{\xu}{\alpha(1-\xu)+\xu}$
$=\alpha c\xb/\xu=\alpha\, p(x)$.
The proof for $n(\alpha\cdot x)$ is analogous.
Property~6: Follows directly by dividing the first and second component of $\alpha\cdot x$.
\hfill$\square$

\begin{lemma}[Distributivity of the scalar multiplication]
\label{lemma:scalardist}
Let $x,y\in\Omega'$. Let $\alpha,\beta \geq 0$.
Then it holds that
\begin{equation}
	\alpha\cdot(x\oplus y)=(\alpha\cdot x)\oplus(\alpha\cdot y)
	\quad\mbox{ and }\quad
	(\alpha+\beta)\cdot x=(\alpha\cdot x)\oplus (\beta\cdot x).
\end{equation}
\end{lemma}
\underline{\it Proof:}
We make extensive use of $\alpha\cdot x\propto(\alpha \xb,\alpha\xd,\xu)$.
First part: On the one hand,
$\alpha\cdot(x\oplus y)\propto
(\alpha[\xu\yb+\yu\xb],\alpha[\xu\yd+\yu\xd],\xu\yu)$.
We have not written the normalization factor; it is not necessary since we already know that
the result is normalized.
On the other hand, $(\alpha\cdot x)\oplus(\alpha\cdot y)\propto$
$(\xu[\alpha \yb]+\yu[\alpha \xb],\xu[\alpha \yd]+\yu[\alpha \xd],\xu\yu)$.

Second part:
On the one hand, $(\alpha+\beta)\cdot x\propto
([\alpha+\beta]\xb,[\alpha+\beta]\xd,\xu)$.
On the other hand, $(\alpha\cdot x)\oplus (\beta\cdot x)$
$\propto ([\beta\cdot x]_{\rm u}[\alpha\cdot x]_{\rm b}+[\alpha\cdot x]_{\rm u}[\beta\cdot x]_{\rm b},
[\beta\cdot x]_{\rm u}[\alpha\cdot x]_{\rm d}+[\alpha\cdot x]_{\rm u}[\beta\cdot x]_{\rm d},
[\alpha\cdot x]_{\rm u}[\beta\cdot x]_{\rm u})$
$\propto(\xu[\alpha\xb]+\xu[\beta\xb],
\xu[\alpha\xd]+\xu[\beta\xd],
\xu^2)$
$\propto (\alpha\xb+\beta\xb,\alpha\xd+\beta\xd,\xu)$.
\hfill$\square$

\begin{lemma}
\label{lemma:multmult}
Let $x\in\Omega'$ and $\alpha,\beta\geq 0$.
Then $\alpha\cdot(\beta\cdot x)=(\alpha\beta)\cdot x$.
\end{lemma}
\underline{\it Proof:}
$\alpha\cdot(\beta\cdot x)\propto\alpha\cdot(\beta\xb,\beta\xd,\xu)$
$\propto(\alpha\beta\xb,\alpha\beta\xd,\xu)$.
\hfill$\square$

\subsection{New discounting rule}
\label{sec:newdiscounting}

We propose a new approach to discounting:
instead of multiplying (part of) the {\em opinions} we multiply the {\em evidence}.
The multiplication is done using our scalar multiplication rule (Def.~\ref{def:mult}).
We return to the example where Alice has an opinion $x\in\Omega'$ about
the trustworthiness of Bob, and Bob has an opinion $y\in\Omega'$ about
some proposition~$P$.
We propose a discounting of the form $g(x)\cdot y$, where $g(x)\geq 0$ is a scalar
that indicates which fraction of Bob's evidence is accepted by Alice.
One can visualize the discounting as a physical transfer of evidence from
Bob to Alice, during which only a fraction $g(x)$ survives, due to
Alice's mistrust and/or uncertainty.
It is desirable to set $g(x)$ in the range $[0,1]$: 
allowing $g(x)<0$ would lead to 
negative {\em amounts} of evidence (not to be confused with the term `negative evidence' which is used for
evidence that contradicts the proposition P); 
allowing $g(x)>1$ would ``amplify''  evidence, i.e., create new evidence out of nothing, which
is clearly unrealistic.

It makes intuitive sense to set 
$\lim_{x\to B}g(x)=1$, $\lim_{x\to D}g(x)=0$ and $g(U)=0$,
or even to set $g(x)=\tilde g(\xb)$, i.e. a function of $\xb$ only,
with $\tilde g(0)=0$ and $\tilde g(1)=1$.
For instance, we could set $g(x)=\xb$.\footnote{Even though taking $g(x)=\xb$ appears to be similar to the $\otimes$ rule (\ref{otimes}), 
we show in Section~\ref{sec:experiments} that they have a very different behavior.}
On the other hand, it could also make sense to set $g(U)>0$, which would represent
the ``benefit of the doubt''.
An intuitive choice would then be $g(x)=x_{\rm b}+ax_{\rm u}$, i.e. the expectation value
corresponding to~$x$.
We postpone the precise details of how the function $g$ can/should be chosen, and
introduce a very broad definition.

\begin{definition}[New generic discounting rule $\boxtimes$]
\label{def:boxtimes}
Let $x,y\in\Omega'$. Let $g: \Omega'\to[0,1]$ be a function.
We define the operation $\boxtimes$ as
\begin{equation}
	x\boxtimes y\stackrel{\rm def}{=}g(x)\cdot y
	=\frac{(\; g(x)\yb,\, g(x)\yd,\,\yu \;)}{(\yb+\yd)g(x)+\yu}
\label{defbox}
\end{equation}
with the $\cdot$ operation as specified in Def.~\ref{def:mult}.
\end{definition}
Differently from $\otimes$, the operator $\boxtimes$ has a well-defined interpretation in terms of evidence handling.
The following theorem states that the evidence underlying $x\boxtimes y$ is a fraction of the evidence underlying $y$ defined by a scalar weight depending on $x$.

\begin{theorem}
\label{th:boxprop}
Let $x,y\in\Omega'$. The operation $\boxtimes$ (Def.~\ref{def:boxtimes})
has the following properties:
\begin{enumerate}
\item
$x\boxtimes y\in\Omega'$.
\item
$p(x\boxtimes y)=g(x)p(y)$ and $n(x\boxtimes y)=g(x)n(y)$.
\item
$(x\boxtimes y)_{\rm b}/(x\boxtimes y)_{\rm d}=\yb/\yd$.
\item
$x\boxtimes U=U$.
\item
Discounting cannot decrease uncertainty, i.e. $(x\boxtimes y)_{\rm u}\geq \yu$.
\end{enumerate}
\end{theorem}
\underline{\it Proof:}
Property~1 follows from $x\boxtimes y=g(x)\cdot y$ and the first property in Lemma~\ref{lemma:scalarprop}.
Property~2: We compute $p(x\boxtimes y)=(x\boxtimes y)_{\rm b}/(x\boxtimes y)_{\rm u}$
using the definition (\ref{defbox}), which yields $g(x)\yb/\yu=g(x)p(y)$. 
For $n(x\boxtimes y)$ the derivation is analogous.
Property~3: Follows directly by dividing the belief and disbelief part of (\ref{defbox}).
Property~4: Follows by setting $\yb=0$, $\yd=0$ and $\yu=1$ in (\ref{defbox}).
Property~5: We have $(x\boxtimes y)_{\rm u}=\frac{\yu}{(\yb+\yd)g(x)+\yu}$.
Since $\yb+\yd+\yu=1$ and $g(x)\in[0,1]$, the denominator of the fraction lies in the range $[\yu,1]$.
\hfill$\square$


\begin{corollary}
\label{corol:BDU}
Let $x,y\in\Omega'$ with $g(x)>0$.
Let $\lim_{y\to B}g(y)=1$, $\lim_{y\to D}g(y)=0$ and $g(U)=0$.
Then the extreme points $B$, $D$, $U$ have the following behavior
with respect to the new discounting rule~$\boxtimes$,
\begin{eqnarray}
	\lim_{y\to B}y\boxtimes x=x \quad&&\quad\quad \lim_{y\to B}x\boxtimes y=B
	\nonumber\\
	\lim_{y\to D}y\boxtimes x=U \quad&&\quad\quad \lim_{y\to D}x\boxtimes y=D
	\nonumber\\
	U\boxtimes x=U. \quad&&
\end{eqnarray}
\end{corollary}
We stress again that the whole `dogmatic' line between $B$ and $D$ is not part of
the opinion space~$\Omega'$, so that we avoid having to deal with infinite amounts of evidence.

\begin{theorem}
\label{th:neversame}
There is no function $g:\Omega'\to[0,1]$ such that $x\boxtimes y=x\otimes y$
for all $x,y\in\Omega'$.
\end{theorem}
\underline{\it Proof:}
On the one hand, we have $x\otimes y=(\xb\yb,\xb\yd,1-\xb\yb-\xb\yd)$.
On the other hand, $x\boxtimes y=\frac{(g(x)\yb,g(x)\yd,\yu)}{g(x)(\yb+\yd)+\yu}$.
Demanding that they are equal yields, after some rewriting,
$g(x)=\xb\yu/[1-\xb(1-\yu)]$. This requires $g(x)$, which is a function of $x$ only,
to be a function of $\yu$ as well.
\hfill$\square$

Being based on the scalar multiplication rule (and hence ultimately on the $\oplus$ rule),
our operation $\boxtimes$
has several properties that $\otimes$ lacks:
(i) right-distribution;
(ii) permutation symmetry of parties that transfer evidence.
This is demonstrated below.

\begin{lemma}
\label{lemma:boxrightdist}
Let $x,y,z\in\Omega'$. Then 
\begin{equation}
	x\boxtimes(y\oplus z)=(x\boxtimes y)\oplus(x\boxtimes z).
\label{boxrightdist}
\end{equation}
\end{lemma}
\underline{\it Proof:}
It follows trivially from $x\boxtimes(y\oplus z)=g(x)\cdot(y\oplus z)$ and
Lemma~\ref{lemma:scalardist}.
\hfill$\square$

This distributive property resolves the issue discussed in Section~\ref{sec:flaw}:
using the $\boxtimes$ operator,
{\em it does not matter if $y$ and $z$ are combined before or after the discounting}.
This  solves the inconsistency caused by the $\otimes$ operation.

Notice also that the left-hand side of (\ref{boxrightdist})
obviously is not double-counting~$x$; hence also the
expression on the right-hand side does not double-count~$x$.
In contrast, the right-hand side expression with $\otimes$
instead of $\boxtimes$ would be double-counting.
We come back to this point in Section~\ref{sec:nodouble}.

\begin{lemma}
\label{lemma:permute}
Let $y,x_1,x_2\in\Omega'$. Then
\begin{equation}
	x_1\boxtimes(x_2\boxtimes y) = x_2\boxtimes(x_1\boxtimes y).
\end{equation}
\end{lemma}
\underline{\it Proof:}
$x_1\boxtimes(x_2\boxtimes y) =g(x_1)\cdot(g(x_2)\cdot y)$.
Using Lemma~\ref{lemma:multmult} this reduces to
$(g(x_1)g(x_2))\cdot y$.
Exactly the same reduction applies to $x_2\boxtimes(x_1\boxtimes y)$.
\hfill$\square$

Lemma~\ref{lemma:permute} generalizes to chains of
discounting:
\begin{equation}
	x_1\boxtimes (x_2\boxtimes(\cdots(x_N\boxtimes y)))
	= [\prod_{i=1}^N g(x_i)]\cdot y.
\label{boxtimeschain}
\end{equation}
Expression (\ref{boxtimeschain}) is invariant under permutation of the
opinions $x_1,\ldots,x_N$.

Note that one property of the $\otimes$ rule is
{\em not} generically present in the $\boxtimes$ rule: associativity.
Whereas the old rule has $x\otimes(y\otimes z)=(x\otimes y)\otimes z$,
the new rule has
\begin{eqnarray}
	x\boxtimes(y\boxtimes z) &= &[g(x)g(y)]\cdot z
	\quad\mbox{ versus }
	\nonumber\\
	(x\boxtimes y)\boxtimes z &=& (g(x\boxtimes y)) \cdot z
	=[g(g(x)\cdot y)]\cdot z.
\label{notassoc}
\end{eqnarray}
However, it is important to realize that {\em the lack of associativity
is not a problem}.
The transfer of evidence along a chain has a very clear ordering,
which determines the order in which the $\boxtimes$ operations have to
be performed. (See Section~\ref{sec:SLflow}.)

Also note that
$\boxtimes$ does not have a left-distribution property
for arbitrarily chosen~$g$.
It takes some effort to define a reasonable function $g$
that yields left-distributivity.



\begin{lemma}
\label{lemma:noleftdist}
There is no function $g:\Omega'\to[0,1]$ that
satisfies $\lim_{s\to B}g(s)=1$
and gives
$(x\oplus y)\boxtimes z=(x\boxtimes z)\oplus(y\boxtimes z)$
for all $x,y,z\in\Omega'$.
\end{lemma}
\underline{\it Proof:}
We consider the limit $x\to B,y\to B$.
On the one hand, $(x\oplus y)\boxtimes z\to B\boxtimes z=z$.
On the other hand,
$(x\boxtimes z)\oplus(y\boxtimes z)\to z\oplus z$.
\hfill$\square$

It may look surprising that we cannot achieve left-distributivity with a
function~$g$ chosen from a very large function space with only a single constraint.
(And a very reasonable-looking constraint at that).
But left-distributivity requires $g(x\oplus y)=g(x)+g(y)$, which
conflicts with the constraint
$\lim_{s\to B} g(s)=1$.

\subsection{New specific discounting rule}
\label{sec:odot}

One way to satisfy $g(x\oplus y)=g(x)+g(y)$
is by setting $g(x)\propto p(x)$.
This approach, however, causes some complications.
Suppose we define $g(x)=p(x)/\theta$, where $\theta$ is some constant.
If the amount of positive evidence ever exceeds $\theta$, then the discounting factor becomes larger
than~1, i.e. amplification instead of reduction, which is an undesirable property.
If we redefine $g$ such that factors larger than~$1$ are mapped back to~1, then we lose
the distribution property.
We conclude that the ``$g$ proportional to evidence'' approach
can only work if the maximum achievable amount of positive evidence
in a given trust network can be upper-bounded by $\theta$.

\begin{definition}[New specific discounting rule $\odot$]
\label{def:odot}
Let $x,y\in\Omega'$.
Let $\theta$ be a threshold larger than the maximum amount of positive evidence
in any opinion that is used for discounting.
We define the operation $\odot$ as
\begin{equation}
	x\odot y \stackrel{\rm def}{=} (\frac{c}{\theta}\; \frac{\xb}{\xu})\cdot y
	= \frac{p(x)}{\theta}\cdot y
\label{odotrule}
\end{equation}
with the $\cdot$ operation as specified in Def.~\ref{def:mult}.
\end{definition}
We stress again that $\theta$ 
depends on the interactions between entities within the system, i.e.\ 
on the structure of the trust network and the maximum amount of positive evidence in the network.

\begin{lemma}[Left-distributivity of $\odot$]
\label{lemma:odotleft}
Let $x,y,z\in\Omega'$. Then
\begin{equation}
	(x\oplus y)\odot z = (x\odot z)\oplus(y\odot z).
\label{odotleft}
\end{equation}
\end{lemma}
\underline{\it Proof:}
The right-hand side evaluates to
$(x\odot z)\oplus(y\odot z)=[g(x)\cdot z]\oplus[g(y)\cdot z]$
$=[g(x)+g(y)]\cdot z$.
The $g$-function in Def.~\ref{def:odot} equals positive evidence divided by $\theta$;
hence $g(x)+g(y)=[p(x)+p(y)]/\theta$.
The left-hand side evaluates to $g(x\oplus y)\cdot z$
with $g(x\oplus y)=\theta^{-1}p(x\oplus y)$
$=[p(x)+p(y)]/\theta$.
\hfill$\square$

\begin{lemma}[Associativity of $\odot$]
\label{lemma:odotassoc}
Let $x,y,z\in\Omega'$. Then
\begin{equation}
	x\odot(y\odot z) = (x\odot y)\odot z .
\label{odotassoc}
\end{equation}
\end{lemma}
\underline{\it Proof:}
From (\ref{notassoc}) we see that $x\odot(y\odot z) =[g(x)g(y)]\cdot z$,
but now we have $g(x)g(y)=\theta^{-2}p(x)\,p(y)$.
From (\ref{notassoc}) we also see
$(x\odot y)\odot z=g(g(x)\cdot y)\cdot z$, but now we have
$g(g(x)\cdot y)=\theta^{-1}p(\theta^{-1}p(x)\cdot y)$
$=\theta^{-2}p(x)\,p(y)$.
\hfill$\square$

\vskip2mm

{\em We are not claiming that $\odot$ is the proper discounting operation to use.}
It has the unpleasant property that the negative evidence underlying
$x$ is completely ignored in the computation of $x\odot y$.
A quick-fix of the form $g(x)\propto p(x)-n(x)$
does not work since it can cause $g(x)<0$ and therefore
$x\boxtimes y\notin\Omega'$.

We note that there is no alternative $g$-function to the
ones discussed above if linearity of $g$ is required.
This is formalized in the following lemma.
\begin{lemma}
\label{lemma:linearg}
The property $g(x\oplus y)=g(x)+g(y)$ can only be achieved by setting
$g(x)=\alpha p(x)+\beta n(x)$, where $\alpha$ and $\beta$ are scalars. 
\end{lemma}
The proof is given in the Appendix.
Note that, for sufficiently smooth $g$, it is possible to prove that the property $g(x\oplus y)=g(x)+g(y)$
implies $g(g(x)\cdot y)=g(x)g(y)$, i.e. associativity.

\subsection{The new discounting rule avoids double-counting}
\label{sec:nodouble}

In Section~\ref{sec:prelimsubjective} we saw that the
consensus operation $\oplus$ should be applied only
to opinions that are derived from {\em independent} evidence.
If this restriction is not obeyed then we speak of double-counting.
Example~\ref{ex:oplustransport} showed that in the SL expression $(x\otimes y)\oplus(x\otimes z)$,
the evidence in $x$ is double-counted.
If we look at the equivalent expression in which $\otimes$ is replaced with $\boxtimes$, we get
\begin{equation}
	(x\boxtimes y)\oplus(x\boxtimes z)= x\boxtimes(y\oplus z).
\label{nodouble1}
\end{equation}
Here there is obviously no double-counting.
Next we look at more complicated EBSL expressions.
\begin{lemma}
\label{lemma:nodouble}
Let $x,y\in\Omega'$ be independent opinions. 
Let $Q,W\in\Omega'$ be opinions independent of $x$ and $y$, but with mutual dependence. 
Then, the expression
\begin{equation}
	(Q\boxtimes x) \oplus (W\boxtimes y)
\label{lemnotdouble}
\end{equation}
does not double-count any of the evidence underlying $Q$ and $W$.
\end{lemma}
\underline{\it Proof:}
The evidence underlying $Q\boxtimes x$
is the evidence from $x$, scalar-multiplied by $g(Q)$.
Likewise, the evidence underlying $W\boxtimes y$
is a scalar multiple of $(p(y),n(y))$.
Since $x$ and $y$ are independent, the evidence
on the left and right side of the `$\oplus$' in (\ref{lemnotdouble}) is
independent.
\hfill$\square$




Based on the result above we can conclude that:
\begin{corollary}\label{cor:double-counting}
Transporting different pieces of evidence over the same link $x$ with the $\boxtimes$ operation is {\em not} double-counting $x$.
\end{corollary}
Thus, many expressions that are problematic in SL become perfectly
acceptable in EBSL, simply because $\boxtimes$ is just an
(attenuating) evidence transport operation,
whereas SL's $\otimes$ is a very complicated thing
that mixes evidence from its left and right operand (Eqs.~\ref{pxy} and~\ref{nxy}).




\section{Flow-based reputation with uncertainty}
\label{sec:SLflow}

In this section we will use the discounting rule $\boxtimes$
without specifying the function~$g$.
We show that EBSL can be applied to arbitrarily connected trust networks
and that the simple recursive approach (\ref{eq:flow+sub}),
with $\otimes$ replaced by $\boxtimes$,
yields consistent results that avoid the double-counting problem.

\subsection{Recursive solutions using EBSL}
\label{sec:disambiguating}

We show that the trust networks 
discussed in Section~\ref{sec:flowSL}, which are problematic in SL, 
can be handled in EBSL. 
We take the EBSL equivalent of the recursive approach in SL
(\ref{eq:functional+sub}), (\ref{eq:flow+sub}), namely
\begin{eqnarray}
	F_{iP} &=& T_{iP} \oplus \bigoplus_{j: j \neq i} (R_{ij} \boxtimes T_{jP})
	\nonumber\\
	R_{ij} &=& A_{ij} \oplus \bigoplus_{k: k \neq i} (R_{ik} \boxtimes A_{kj})
	\quad\quad\mbox{ for }i\neq j,
\label{EBSLrecursion}
\end{eqnarray}
and demonstrate that these equations yield acceptable 
results in the case of the trust networks depicted in
Figures~\ref{fig:problem1} and \ref{fig:loop123}.

For Figure~\ref{fig:problem1} we obtain the EBSL equivalent of (\ref{F1P}) 
by using (\ref{EBSLrecursion}) recursively
as follows,

$
\begin{array}{ll}
	F_{1P}= & R_{17}\boxtimes T_{7P}\\
	R_{17}= & R_{16}\boxtimes A_{67} \\
	R_{16}=& (R_{14}\boxtimes A_{46})\oplus (R_{15}\boxtimes A_{56}) \\
	R_{15}=& (R_{14}\boxtimes A_{45})\oplus (R_{13}\boxtimes A_{35})\\	
	R_{14}=& R_{13}\boxtimes A_{34}\\
	R_{13}=& R_{12}\boxtimes A_{23}\\
	R_{12}=&  A_{12}.
\end{array}
$
\medskip

\noindent
By substituting these expressions from bottom to top
we get the end result for $F_{1P}$.
The result is very similar to (\ref{F1P}), but now we have lots of
brackets because $\boxtimes$ is not associative.
We inspect $R_{16}$,
\begin{eqnarray}
\label{R16EBSL}
	R_{16}&=&((A_{12}\smallbox A_{23})\smallbox A_{34})\smallbox A_{46} \oplus 
	\\ &&
	\left\{
	((A_{12}\smallbox A_{23})\smallbox A_{34})\smallbox A_{45} \oplus
	(A_{12}\smallbox A_{23})\smallbox A_{35}
	\vphantom{M^{M^M}}\right\}\smallbox A_{56}.
	\nonumber
\end{eqnarray}
We observe that the links $A_{12}$ and $A_{23}$ occur three times.
The link $A_{34}$ occurs twice.
However, the computation of $R_{16}$ only relies on the evidence in 
$A_{46}$ and $A_{56}$; all the other opinions $A_{ij}$ serve as `transport',
i.e. merely providing weights multiplying the evidence in $A_{46}$ and $A_{56}$.
Therefore, there is no double counting of evidence.

In the case of Figure~\ref{fig:loop123}, i.e. with a loop,
recursive use of (\ref{EBSLrecursion}) gives 
the direct EBSL equivalent of (\ref{loopSL}). We have
\begin{eqnarray}
	R_{13}&=& R_{12}\boxtimes A_{23}
	\\
	R_{12}&=& A_{12}\oplus R_{13}\boxtimes A_{32}
	\nonumber\\ &=&
	A_{12}\oplus (R_{12}\boxtimes A_{23})\boxtimes A_{32}.
\label{loopy1}
\end{eqnarray}
Eq.\,(\ref{loopy1}) gives an expression for the unknown $R_{12}$
that contains $R_{12}$.
It can be solved in two ways.
The first is to repeatedly substitute (\ref{loopy1}) into itself.
We define a mapping $f(x)=A_{12}\oplus (x\boxtimes A_{23})\boxtimes A_{32}$.
From (\ref{loopy1}) we see that $R_{12}$ is a fixed point of~$f$.
The fixed point can be found approximately by setting $x=A_{12}$
and repeatedly applying~$f$ until the output does not change any more.
\begin{equation}
	R_{12}\approx f^r(A_{12}) \quad\quad\mbox{ for large }r.
\end{equation}
\begin{equation}
	R_{12}=A_{12}\oplus ([A_{12}\oplus (\{A_{12}\oplus\cdots\}\smallbox A_{23})\smallbox A_{32}]
	\smallbox A_{23})\smallbox A_{32}.
\label{loopy2}
\end{equation}
Convergence will be fast if $A_{23}$ and $A_{32}$ contain a lot of uncertainty.
In Eq.\,(\ref{loopy2}) we observe that
(i)
The evidence contributing to $R_{12}$ comes from the first $A_{12}$ 
and from the final~$A_{32}$.
(ii)
All the other occurrences of opinions $A_{ij}$ are only used to compute the weights
for the scalar multiplication.

The second method is to
treat (\ref{loopy1}) as two independent equations 
in two unknowns (the two independent components
of $R_{12}\in\Omega'$) and to solve them algebraically.
This can be quite difficult if the function $g$ is complicated, since $g$ has to be applied twice,
\begin{eqnarray}
	&&R_{12}= A_{12}\oplus g(g(R_{12})\cdot A_{23})\cdot A_{32}
	\\
	&&\Leftrightarrow \nonumber\\
	&&\left\{\matrix{
	p(R_{12})=p(A_{12})+ g(g(R_{12})\cdot A_{23}) p(A_{32})
	\cr
	n(R_{12})=n(A_{12})+ g(g(R_{12})\cdot A_{23}) n(A_{32}).
	}
	\right.
\label{EBSLggR12}
\end{eqnarray}
The solution is simple in the following special cases:
\begin{itemize}
\item
If $g(A_{12})=0$ and $g(U)=0$ then $R_{12}=A_{12}$.
\item
If $A_{23}\neq U$, $g(x)>0$ for $x\neq U$, and $A_{32}\to B$ then $R_{12}\to B$.
The direct $A_{12}$ can be overwhelmed by the indirect $A_{32}$, even
if user $1$ has little trust in user~$2$. 
This demonstrates the danger of allowing opinions close to full belief. 
\item
If
$A_{23}\to B$ and $g(B)=1$ then $R_{12}\to A_{12}\oplus A_{32}$.
\end{itemize}
 
\subsection{Recursive solution in matrix form}
\label{sec:matrixform}

The recursive equation (\ref{EBSLrecursion})
for obtaining the $R_{ij}$ solutions
can be rewritten in matrix notation.
We choose $g$ such that $g(U)=0$.
We are looking for the off-diagonal components of a matrix $R$ or,
equivalently, for a complete matrix $R$ which has the uncertainty `$U$' on its diagonal.
Let $X$ be an $n\times n$ matrix containing opinions; it is allowed to have a non-empty diagonal,
e.g. $X_{ii}\neq U$.
We define a function $f$ as 
\begin{equation}
\label{eq:fix-point}
	f(X)\stackrel{\rm def}{=}A\oplus \{[{\rm offdiag}\, X]\boxtimes A\},
\end{equation}
where `${\rm offdiag}\, X$' is defined as $X$ with its diagonal replaced by $U$ entries, and
an expression of the form $X\boxtimes A$ is a matrix defined as
$(X\boxtimes A)_{ij}=\bigoplus_{k}X_{ik}\boxtimes A_{kj}$.
Note that $f(X)$ is a matrix that can have a non-empty diagonal.
Solving (\ref{EBSLrecursion}) is equivalent\footnote{
An even more compact formulation is possible if one is willing to temporarily use the full belief $B$
in the computations. Let `$\bf 1$' be the diagonal matrix. Let $Z=B{\bf 1}+R$.
Solving (\ref{EBSLrecursion}) for $R$ is equivalent to solving $Z=B{\bf 1}\oplus (Z\boxtimes A)$
for $Z$.
}
to the following procedure:
\begin{enumerate}
\item
Find the fixed point $X_*$ satisfying $f(X_*)=X_*$.
\item
Take $R={\rm offdiag}\, X_*$.
\end{enumerate}
One approach to determine the fixed point $X_*$ is to pick
a suitable starting matrix $X_0$ and then keep applying $f$ until the output does not change any more,
$X_*\approx f^N(X_0)$.
Another approach is to treat the formula $f(X_*)=X_*$ as a set of algebraic equations,
whose complexity is affected by the choice of the $g$ function.

At this point, two important questions have to be answered: \emph{i)} whether the recursive approach for 
(\ref{eq:fix-point}) converges, i.e.\ if the fixed point exists, and \emph{ii)}, when it
converges, whether the fixed point solution is unique.
When there are no loops in the network, then trivially we have convergence and the fixed point is unique.
Intuitively, the repeated applications of $f$ after the trust network has been completely explored do not propagate additional evidence.

In the case of general networks the situation is more complicated.
We can prove that there is {\em no divergence}.
In every iteration of the mapping $X\mapsto f(X)$, 
the new value of $X$ is of the form
$X_{ij}=A_{ij}\oplus\bigoplus_k g(({\rm offdiag}\, X)_{ik})\cdot A_{kj}$.
We see that the evidence in each $A_{kj}$ gets multiplied by a scalar smaller than~$1$.
Hence, no matter how many iterations are done, the amount of evidence
about user $j$ that is contained in $X$ can never exceed the amount of evidence about $j$ present in~$A$.
This puts a hard upper bound on the amount of evidence in ${\rm offdiag}\, X$,
which prevents the solution from `running off' towards full Belief.
Hence, the evidence underlying $R_{ij}$ cannot be greater of the total amount of evidence underlying the opinions in $A$ about user $j$.

It can be observed that, being flow-based, our fixed point equation for the matrix $R$ 
has the same form as the fixed point equation for a Markov Chain.
The main difference is that in an ordinary Markov chain there is a real-valued
transition matrix whereas we have opinions
$A_{ij}\in\Omega^*$, and in our case multiplication of reals is replaced by $\boxtimes$
and addition by~$\oplus$.
In spite of these differences, we observe in our experiments that
every type of behavior of Markov Chain flow also occurs
for~$R$.
Indeed, experiments on 
real data show that we indeed have convergence (see Section~\ref{sec:experiments}).
Moreover, for some fine-tuned instances of the $A$-matrix, which are exceedingly unlikely to occur naturally,
oscillations can exist just like in Markov chains; after a number of iterations the powers of $A$ jump back and forth
between two states. Just as in flow-based reputation (\ref{reputationrA}), 
adding the direct opinion matrix $A$ in each iteration dampens the oscillations and causes convergence.

\subsection{Recursive solutions using the $\odot$ discounting rule}
\label{sec:special}

We investigate what happens when we replace the generic EBSL discounting operation $\boxtimes$
by the special choice $\odot$ as specified in Def.\,\ref{def:odot}.
First we consider the case of Figure~\ref{fig:problem1}. 
The generic Eq.\,(\ref{R16EBSL}) reduces to
\begin{eqnarray}
	R_{16}&=& 
	A_{12}\odot A_{23}\odot A_{34}\odot A_{46}\; \oplus
	\nonumber\\ && 
	A_{12}\odot A_{23}\odot A_{35}\odot A_{56}\; \oplus
	\nonumber\\ && 
	A_{12}\odot A_{23}\odot A_{34}\odot A_{45}\odot A_{56}
	\nonumber\\ &=&
	A_{12}\odot A_{23}\odot
\label{factor123}
	\left\{
	A_{34}\odot A_{46}\oplus (A_{35}\oplus A_{34}\odot A_{45})\odot A_{56}
	\right\}.
\end{eqnarray}
Notice that we do not have to put brackets around chains of $\odot$ operations
because they are associative (Lemma~\ref{lemma:odotassoc}).
Also notice that in (\ref{factor123})
the common `factor' $A_{12}\odot A_{23}$ has been pulled outside the brackets.
We are allowed to add and remove brackets at will because of the associativity and full distributivity
of $\odot$.

Next we consider Figure~\ref{fig:loop123}, the loop case.
Eq.\,(\ref{EBSLggR12}) reduces to 
\begin{equation}
	\left\{ \matrix{
	p(R_{12})= p(A_{12})+\frac1{\theta^2}p(R_{12})p(A_{23}) p(A_{32}) \cr
	n(R_{12})= n(A_{12})+\frac1{\theta^2}p(R_{12})p(A_{23}) n(A_{32})
	}\right.
\end{equation}
which is easily solved,
\begin{eqnarray}
\label{pR12fraction}
	p(R_{12})&=& \frac{p(A_{12})}{1-\theta^{-2}p(A_{23})p(A_{32}) }
	\\ 
	n(R_{12})&=& n(A_{12})
	+\frac{p(A_{12})p(A_{23})}{\theta^2-p(A_{23})p(A_{32}) }n(A_{32}). 
\end{eqnarray}
We observe that the system parameter $\theta$ has to be chosen with great caution.
If values $p(A_{ij})$ can get too close to $\theta$ then
the fraction in (\ref{pR12fraction}) may explode and may result
in $p(R_{12})>\theta$, which is problematic.
Let us define $p_{\rm max}=\max_{ij}p(A_{ij})$.
Then it is necessary to set $\theta\geq p_{\rm max}\cdot(1+\sqrt 5)/2$.
(This bound is obtained by setting
$p(A_{12})=p(A_{23})=p(A_{32})=p_{\rm max}$ in (\ref{pR12fraction}) 
and demanding that $p(R_{12})\leq\theta$.) 

\section{Evaluation}
\label{sec:experiments}

We have implemented our flow-based reputation model with uncertainty and performed a number of experiments to evaluate the practical applicability and ``accuracy'' of the model using both synthetic and real-life data.
Note that, while it is possible to define some limiting situations (synthetic data) in which a certain result is expected, in general numerical experiments cannot `prove' which approach is right because there is no `ground truth' solution to the reputation problem that we could compare against.
The only thing that can be verified by numerics is: (i) do the results make sense? (ii) is the method practical?
Thus, we have used synthetic data to compare the accuracy of the opinions computed using different reputation models. 
On the other hand, we used real-life data to study the practical applicability.

Experiments for assessing the robustness of the reputation model against attacks like slandering, self-promotion and Sybil attacks \cite{DOUC-02-IPTPS,Hoffman:2009,flowbase}, have not been considered in this work and are left for future work, 
as our goal here is the definition of the mathematical foundation for the development of reputation systems.
A study of the robustness against attacks requires to consider many other aspects that are orthogonal to this work.

In the remainder of this section, first we briefly present the implementation; then we report and discuss the results of the experiments.

\subsection{Implementation}

We have developed a tool 
in Python.
It implements the procedure for computing the fixed point described in Section~\ref{sec:matrixform}.
All SL and EBSL computation rules  presented in this paper have been implemented in a Python library.

The tool takes as input a log containing recorded user interactions. 
Based on the evidence contained in the log, the tool extracts the direct referral trust matrix~$A$. 
The tool repeatedly iterates the recursive equation until it converges, that is, the difference between the new matrix $R^{(k+1)}$ and the previous one $R^{(k)}$ is less than a certain threshold.
In particular the termination condition is set as follows:
\begin{equation}
    \sum_{i,j}\delta(R^{(k+1)}_{ij},R^{(k)}_{ij})<10^{-10}
\end{equation}
where $\delta(x,y) \stackrel{\rm def}{=} |x_{\rm b}-y_{\rm b}|+|x_{\rm d}-y_{\rm d}|+|x_{\rm u}-y_{\rm u}|$.

\subsection{Synthetic Data}
\label{sec:numerical-example}

We have conduced a number of experiments  using synthetic data
to analyze and compare the different approaches for trust computation.
The goal of these experiments is to analyze the behavior of the reputation models in a number of limiting situations 
for which it is known a priori how the result should behave.

\paragraph{Experiment Settings}
The experiments are based on the trust network in Figure~\ref{fig:problem1}.
We considered six approaches: 
\emph{(i)} the flow-based method without uncertainty in Eq.~(\ref{reputationrA}); 
\emph{(ii)} the flow-based SL approach presented in Section~\ref{sec:flowSL}; 
\emph{(iii)} SL in which the specification of the trust network is transformed to canonical form by removing the edge from $4$ to $5$ 
(i.e., $A_{45}$ is set to $U$ in Eq.~(\ref{eq:canonical})); 
\emph{(iv)} EBSL with $g(x)=x_b$; 
\emph{(v)} EBSL with $g(x)=\sqrt x_b$; and 
\emph{(vi)} EBSL using the operator $\odot$. 

To make the results comparable, we specify the amount of positive and negative evidence $(p,n)$ for each edge in the trust network and use such evidence to compute the opinions used for EBSL and SL as well as
the aggregate ratings ($A_{xy}$) used for flow-based reputation (without uncertainty).
The mapping between evidence and opinion is computed using (\ref{mappingJosang}).
For the mapping between evidence and aggregated ratings, we use an approach similar to the one presented in \cite{flowbase}:
\begin{equation}\label{eq:aggregated-ratings}
A_{xy} =\frac{1}{2} + \frac{p-n}{2e}
\end{equation}
where $A_{xy}= 1$ means fully trusted, $A_{xy}=0$ fully distrusted, and $A_{xy}= \frac{1}{2}$ neutral.

For the analysis we consider three cases.
Table~\ref{tab:input} presents the evidence along with the derived opinions and aggregated ratings for our first case ({\bf C1}).
In this case, we use $\theta=1000$ for EBSL using operator $\odot$.
In the second case ({\bf C2}), we consider the same evidence except for the edge from $7$ to $P$ which is now $(10,900)$.
The corresponding opinion and aggregate rating are $T_{7P}=(0.011,0.987,0.002)$ and $A_{7P}=0.011$ respectively.
In the last case ({\bf C3}), we consider the evidence for all edges to be $(10000,0)$ and 
we set $\theta=20000$.
In this case, all opinions are equal to $(0.9998,0.0000,0.0002)$ and aggregated ratings are equal to $1$.

\begin{table}[!t]
\centering
\begin{tabular}{ccccc}
Source & Target & Evidence & Opinion & $A_{xy}$\\
\hline
1 & 2 &  $(400,300)$ & $(0.570,0.427,0.003)$ & $0.571$\\
2 & 3 &  $(10,5)$  & $(0.588,0.294,0.118)$ & $0.667$\\
3 & 4 &  $(500,0)$ & $(0.996,0.000,0.004)$ & $1.0$\\
3 & 5 &  $(500,0)$ & $(0.996,0.000,0.004)$ & $1.0$\\
4 & 5 &  $(500,0)$ & $(0.996,0.000,0.004)$ & $1.0$\\
4 & 6 &  $(500,0)$ & $(0.996,0.000,0.004)$ & $1.0$\\
5 & 6 &  $(500,0)$ & $(0.996,0.000,0.004)$ & $1.0$\\
6 & 7 &  $(5,5)$   & $(0.417,0.417,0.166)$ & $0.5$\\
7 & P &  $(10,90)$ & $(0.098,0.882,0.020)$ & $0.1$\\
\hline
\end{tabular}
\caption{\it Evidence, opinions, and aggregate ratings for case {\bf C1}.}
\label{tab:input}
\end{table}

%

\paragraph{Results}
The results of the trust computation are presented in Table~\ref{tab:comparison} in terms of opinions (trust value for flow-based method), and in Table~\ref{tab:comparison-evidence} in terms of amount of evidence.
Note that in Table~\ref{tab:comparison-evidence} we have not included the amount of evidence for the flow-based approach of (\ref{reputationrA}) as it it not possible to reconstruct it from trust values.



\begin{table}[!t]
\centering
\begin{tabular}{lcccc}
& {\bf C1}& {\bf C2}& {\bf C3} \\
\hline
Flow-based               & $0.401$               & $0.392$               & $0.501$\\
Flow-SL                  & $(0.024,0.220,0.756)$ & $(0.003,0.246,0.751)$ & $(0.9993,0.0000,0.0007)$  \\
SL (canonical form)      & $(0.014,0.123,0.863)$ & $(0.002,0.137,0.861)$ & $(0.9990,0.0000,0.0010)$    \\
EBSL ($g(x)=x_b$)        & $(0.095,0.859,0.046)$ & $(0.011,0.984,0.005)$ & $(0.9998,0.0000,0.0002)$ \\
EBSL ($g(x)= \sqrt x_b$) & $(0.097,0.873,0.030)$ & $(0.011,0.986,0.003)$ & $(0.9998,0.0000,0.0002)$ \\
EBSL ($\odot$)           & $(0.000,0.000,1.000)$ & $(0.000,0.006,0.994)$ & $(0.9970,0.0000,0.0030)$    \\
\hline
\end{tabular}
\caption{\it Comparison in terms of trust values $r_{1P}$ and opinions $F_{1P}$.}
\label{tab:comparison}
\end{table}




\begin{table}[!t]
\centering
\begin{tabular}{lrrrr}
&\multicolumn{1}{c}{\bf C1}& \multicolumn{1}{c}{\bf C2}& \multicolumn{1}{c}{\bf C3}\\
\hline
Flow-SL                   & $(0.065,0.582)$  & $(0.007,0.655)$   & $(2763.7,0)$  \\
SL (canonical form)       & $(0.032,0.284)$  & $(0.004,0.319)$   & $(1999.16,0)$ \\
EBSL ($g(x)=x_b$)         & $(4.166,37.490)$ & $(4.166,374.901)$ & $(9998,0)$    \\
EBSL ($g(x)=\sqrt x_b$)   & $(6.455,58.095)$ & $(6.455,580.946)$ & $(9999,0)$    \\
EBSL ($\odot$)            & $(0.000,0.001)$  & $(0.000,0.011)$   & $(781.25,0)$  \\
\hline
\end{tabular}
\caption{\it Comparison in terms of amount of evidence underlying opinion $F_{1P}$.}
\label{tab:comparison-evidence}
\end{table}

The results confirm our expectation about the impact of the trust network representation on the trust computation when SL is used.
As expected, the uncertainty component of opinion $F_{1P}$ computed using SL is larger when the trust network in Figure~\ref{fig:problem1} is represented in canonical form.
Indeed, in this case some trust information (i.e., evidence) is discarded (Recall that uncertainty depends on the amount of evidence; the larger the amount of evidence, the lower the uncertainty. See Section~\ref{sec:evidencemap}).
In contrast, the representation of the trust network in Eq.~(\ref{F1P}) is affected by double counting, 
leading to more evidence and thus to a lower uncertainty component of $F_{1P}$.

The results show that the SL and EBSL approaches preserve the ratio between belief and disbelief components (Table~\ref{tab:comparison}) and consequently the ratio between positive and negative evidence (Table~\ref{tab:comparison-evidence}).
This ratio is close to the one between the positive and negative evidence underlying the functional trust $T_{7P}$.
If the amount of evidence increases ({\bf C2}), one would expect that the amount of evidence underlying opinion $F_{1P}$ increases proportionally to the increase of the amount of evidence underlying $T_{7P}$ (Theorem~\ref{th:boxprop}).
Accordingly, the amount of positive evidence underlying $F_{1P}$ should be the same in {\bf C1} and {\bf C2}, and the amount of negative evidence underlying $F_{1P}$ in {\bf C2} should be ten times the amount of negative evidence in {\bf C1}.
We can observe in Table~\ref{tab:comparison-evidence} that this is true for EBSL but not for SL.
This is explained by the fact that $x \otimes y$ is not an $x$-dependent multiple of the evidence underlying~$y$,
 as was shown in Eqs.~(\ref{pxy}) and~(\ref{nxy}). 

Finally, in the last case ({\bf C3}) we have considered a limiting case where every trust relation in the network of Figure~\ref{fig:problem1} is characterized by a large amount of positive evidence.
Here, one would expect that the opinion $F_{1P}$ is close to $(1,0,0)$ and the trust value $r_{1P}$ close to~$1$.
From Table~\ref{tab:comparison} we can observe that SL and all EBSL approaches meet this expectation.
However, if we look closely at the evidence underlying such an opinion (Table~\ref{tab:comparison-evidence}), we can observe that when the SL discounting operator $\otimes$ and EBSL operator $\odot$ are used, a large amount of evidence is ``lost'' on the way.
In contrast, we expect the amount of evidence underlying $F_{1P}$ to be close to that of $T_{7P}$ (Theorem~\ref{th:boxprop}).
Table~\ref{tab:comparison-evidence} shows that EBSL, both for $g(x)=x_b$ and $g(x)=\sqrt x_b$, preserves the amount of evidence when referral trust relations are close to full belief.

Moreover, Table~\ref{tab:comparison} shows that the value of $r_{1P}$ is close to neutral trust rather than to full trust.
This can be explained by Eq.~(\ref{reputationrA}) and the impossibility to express uncertainty.
On the one hand, at each iteration Eq.~(\ref{reputationrA}) computes a weighted average of aggregated ratings where weights are equal to the trust a user places in the users providing recommendations.
On the other hand, the flow-based approach does not distinguish between neutral trust (equal amount of positive and negative evidence) and full uncertainty (zero evidence).
In particular, the lack of evidence between two users is represented in the matrix of aggregated ratings $A$ as neutral trust (see Eq.~\ref{eq:aggregated-ratings}).
In sparse trust networks (i.e., networks with only a few edges) like the one in Figure~\ref{fig:problem1}, the neutral trust used to express uncertainty has a significant impact on the weighted average used to compute trust values.\footnote{It worth noting that a trust value close to $1$ can be obtained by Eq.~\ref{reputationrA} only if the trust network is a complete graph in which the aggregate rating associated with each edge is close to $1$.}
These results demonstrate that the ability to express uncertainty is fundamental to capture the actual trustworthiness of a target, which is one of the main motivations for this work.

\subsection{Real-life Data}

We performed a number of experiments using real-life data to assess the practical applicability of EBSL 
and flow-based reputation models built on top of EBSL.
In particular, we study the impact of various discounting operators on the propagation of evidence and the convergence speed of the iterative procedure.

\paragraph{Experiment Settings}
For the experiments we used a dataset derived from a BitTorrent-based client called Tribler \cite{Pouwelse:2008}. 
The dataset consists of information about 10,364 nodes and 44,796 interactions between the nodes.
Each interaction describes the amount of transferred data in bytes from a node to another node.
The amount of transferred data can be either negative, indicating an upload from the
first node to the second node, or positive, indicating a download.

To provide an incentive for sharing information, some BitTorrent systems require users to have at least a certain ratio of uploaded vs. downloaded data.
Along this line, we treat the amount of data uploaded by a user as positive evidence, and the downloaded amount as negative evidence.
Intuitively, positive evidence indicates a user's inclination to share data and thus to contribute to the community.



It is worth noting that Tribler has a high population turnover and, thus, the dataset contains very few long living and active nodes alongside many loosely connected nodes of low activity \cite{GkorouVPE13}.
This results in a direct referral trust matrix that is very sparse (i.e., most opinions are full uncertainty).
In this sparse form it is inefficient to do large matrix multiplications.
To this end, we have grouped the nodes into 200 clusters, each of which contains about 50~nodes.
Intuitively, a cluster may be seen as the set of nodes under the control of a single user.





For the experiments with real data, we considered four reputation models: 
\emph{(i)} the flow-based SL approach presented in Section~\ref{sec:flowSL};  
\emph{(ii)} EBSL with $g(x)=x_b$; 
\emph{(iii)} EBSL with $g(x)=\sqrt x_b$; and 
\emph{(iv)} EBSL using the operator $\odot$. 
Note that, due to the large number of interconnected loops in the trust network, we did not consider SL in which the trust network is transformed into a canonical form.


\begin{figure}[!t]
\centering
\begin{subfigure}[b]{.48\linewidth}
\includegraphics[width=60mm]{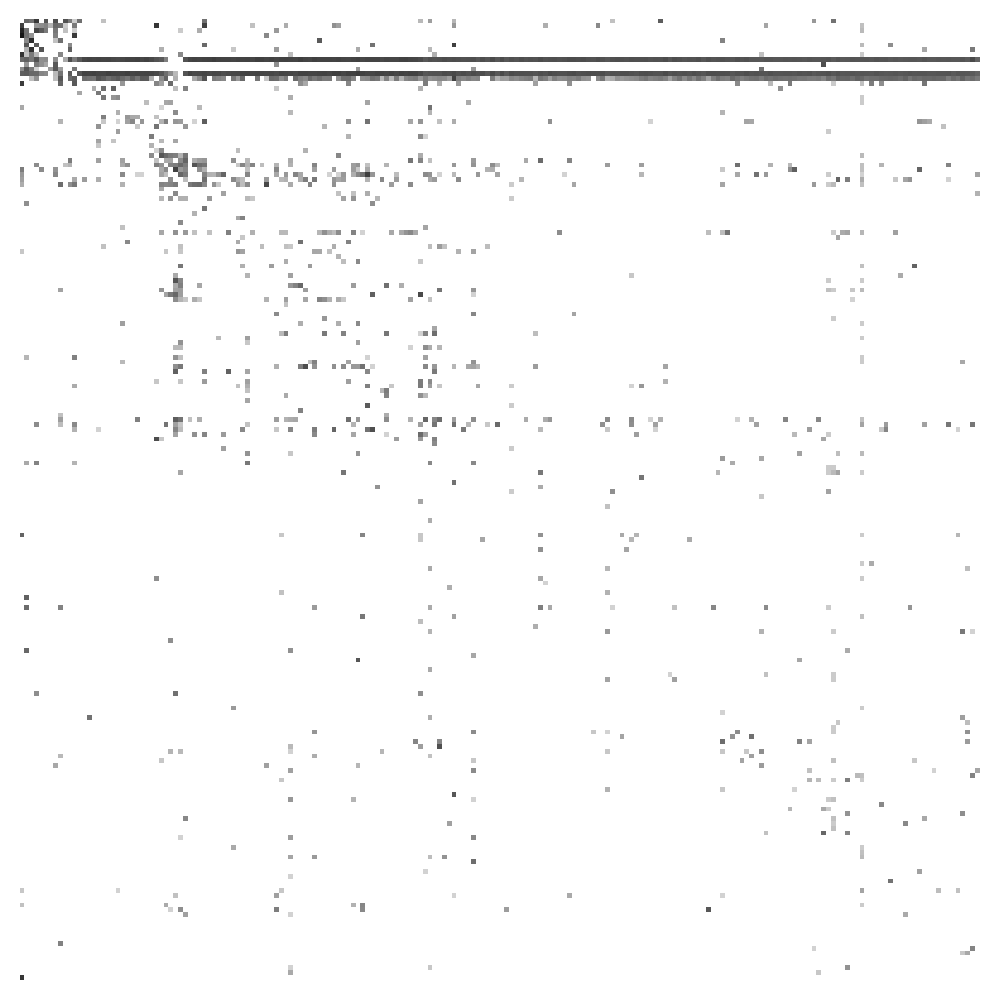}
\caption{Positive evidence}
\label{fig:initial-case1}
\end{subfigure}
~~~
\begin{subfigure}[b]{.48\linewidth}
\includegraphics[width=60mm]{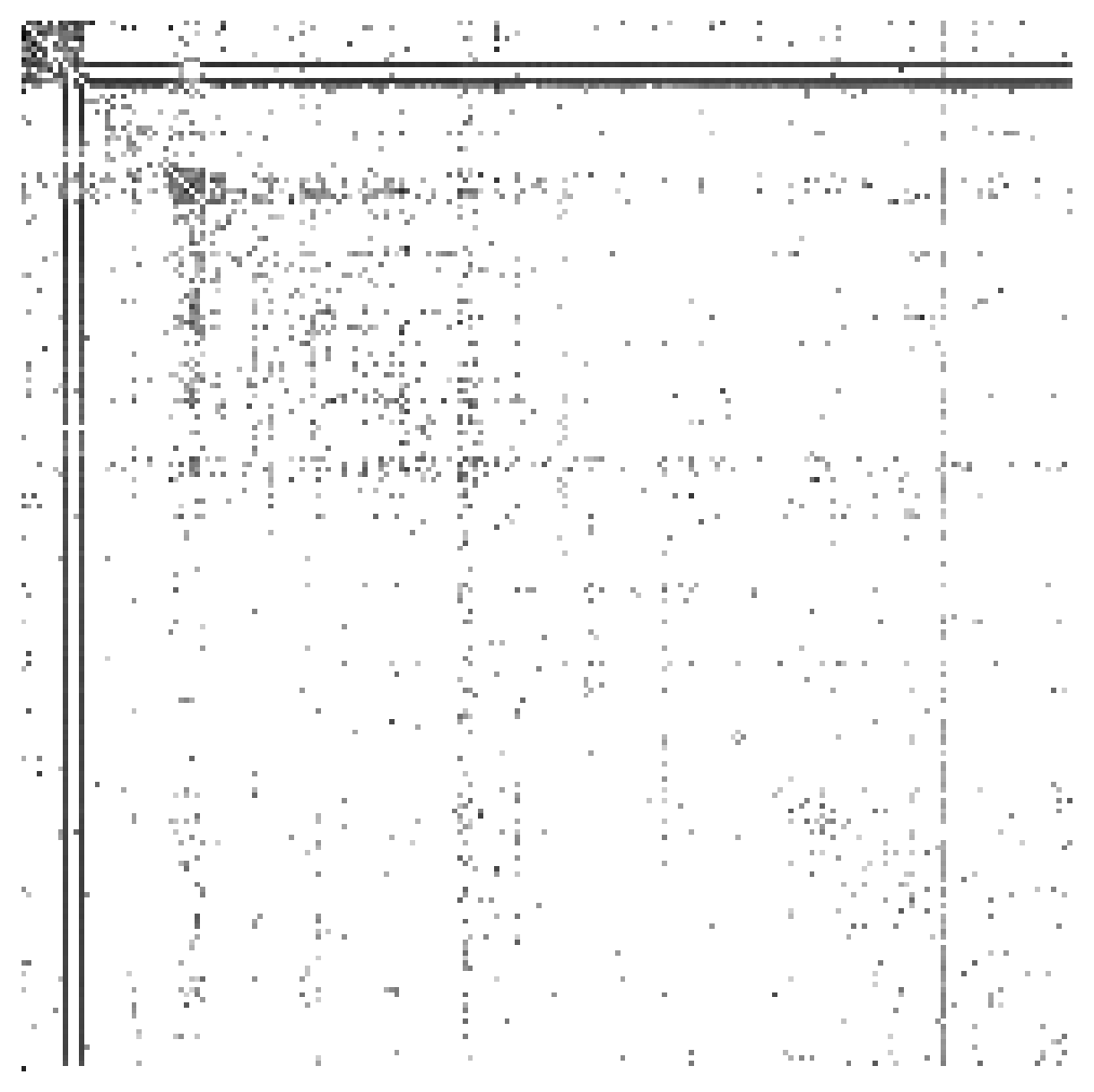}
\caption{Sum of positive and negative evidence}
\label{fig:initial-case2}
\end{subfigure}
\caption{\it The evidence in the direct opinion matrix~$A$ for the Tribler data.
For each pair $(i,j)$ the amount of evidence underlying the opinion of
$i$ about $j$ is shown as a shade of gray, using a logarithmic gray scale.
White corresponds to zero, black to $8.7\cdot10^6$, which is the maximum amount of evidence
occurring in a single matrix entry in the experiments.
}
\label{fig:initial-Tribler}
\end{figure}

In all four models we computed the final referral trust matrix~$R$.
The amount of evidence in the matrix~$A$ is visualized in Figure~\ref{fig:initial-Tribler}.
Figure~\ref{fig:initial-case1} presents the amount of positive evidence, and  Figure~\ref{fig:initial-case2} the total amount of evidence 
(sum of positive and negative).
We can observe the presence of a few active users who had interactions with a lot of other users (visible as dark lines). 
A horizontal dark line in Figure~\ref{fig:initial-case1}
indicates a user who downloaded data from many other users.
The vertical dark lines in Figure~\ref{fig:initial-case2}
represent negative evidence: many users uploading to the same few users.
Note that Figure~\ref{fig:initial-case2} is not symmetric,
since an interaction never results in user feedback from both sides.



It is also interesting to note the clusters of strongly connected users who often interact with each other (for instance the top-left corner).

\paragraph{Results}


\begin{figure}[!t]
\centering
\begin{subfigure}[b]{.48\linewidth}
\includegraphics[width=60mm]{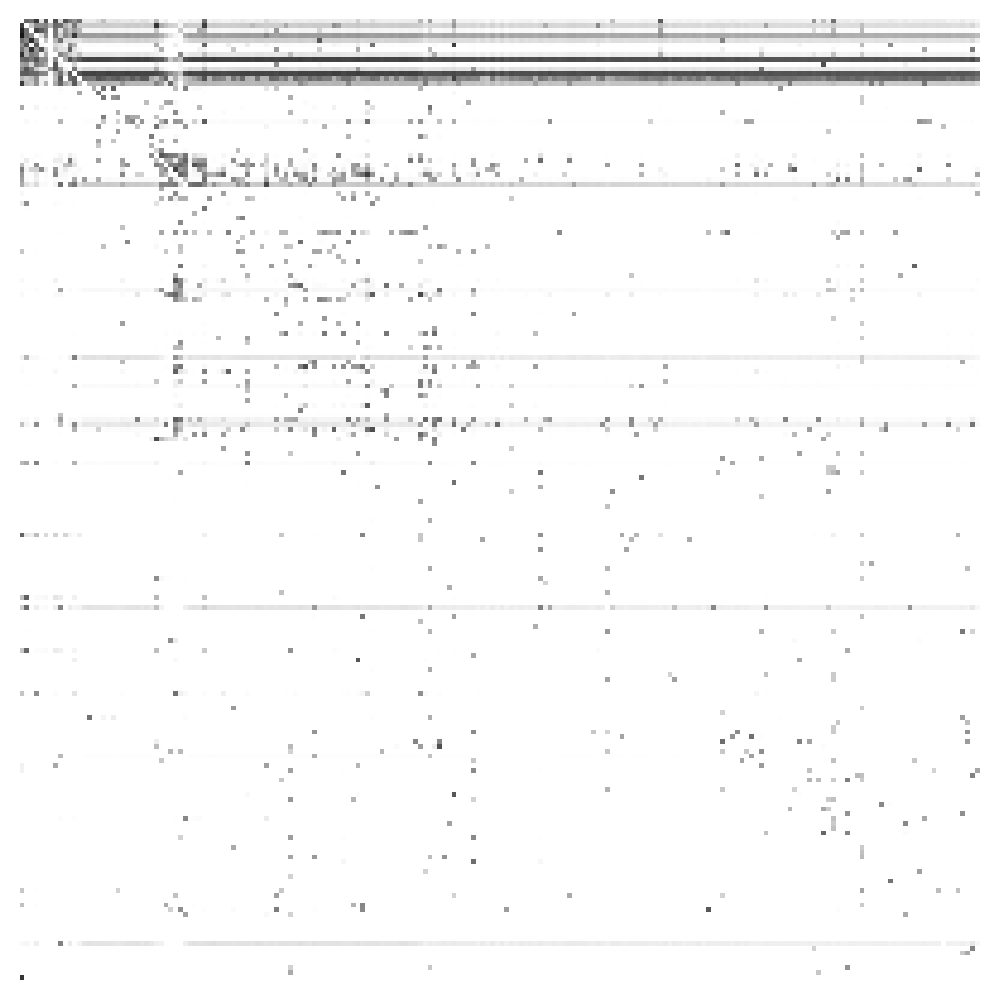}
\caption{$R$ obtained using $\odot$}
\label{fig:odot-case1}
\end{subfigure}
~~~
\begin{subfigure}[b]{.48\linewidth}
\includegraphics[width=60mm]{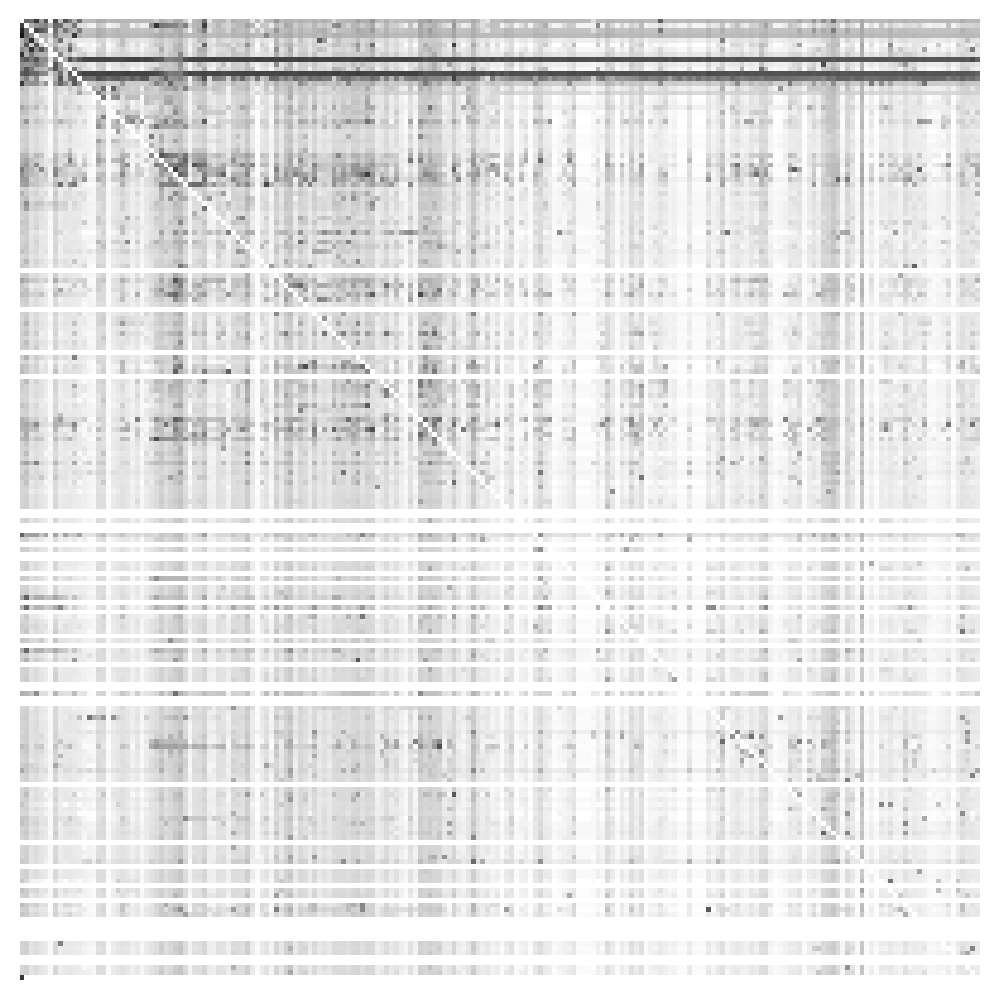}
\caption{$R$ obtained using $\otimes$}
\label{fig:otimes-case1}
\end{subfigure}
\begin{subfigure}[b]{.48\linewidth}
\includegraphics[width=60mm]{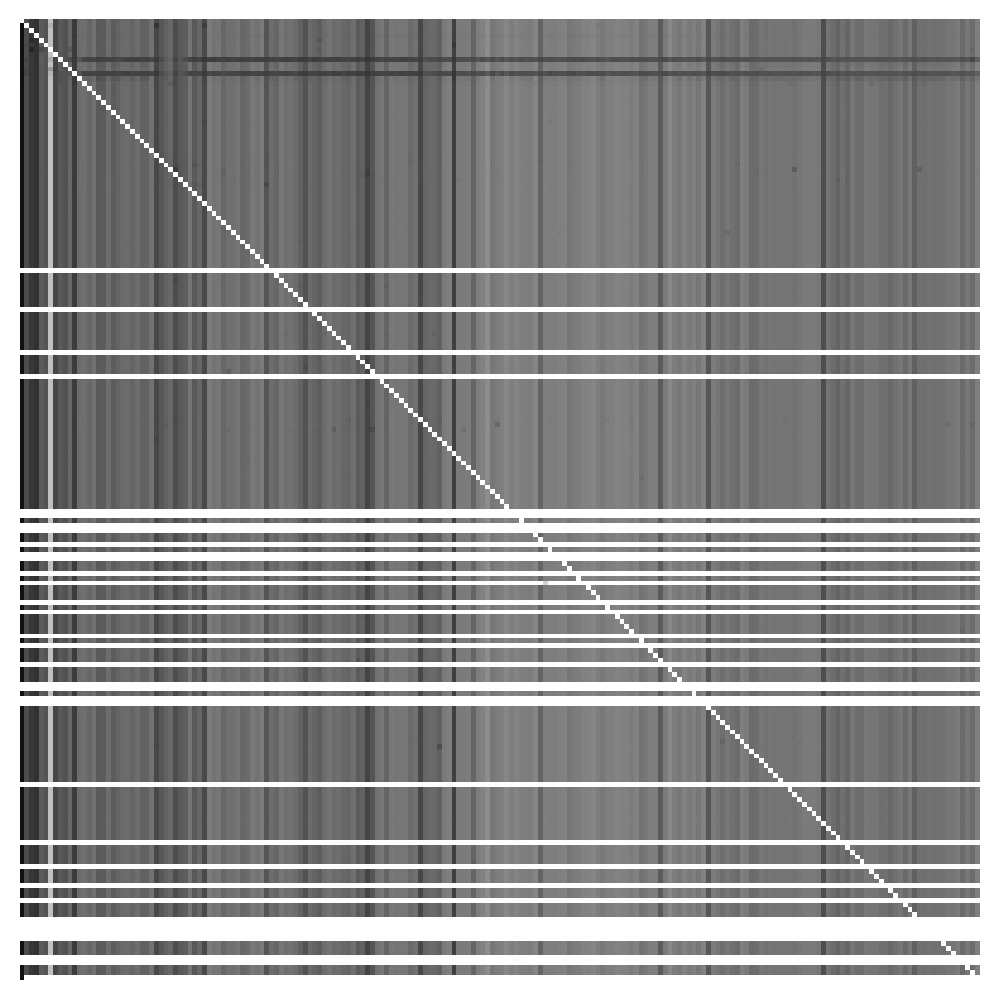}
\caption{$R$ obtained using $\boxtimes$ with $g(x)=x_{\rm b}$}
\label{fig:boxtimexb-case1}
\end{subfigure}
~~~
\begin{subfigure}[b]{.48\linewidth}
\includegraphics[width=60mm]{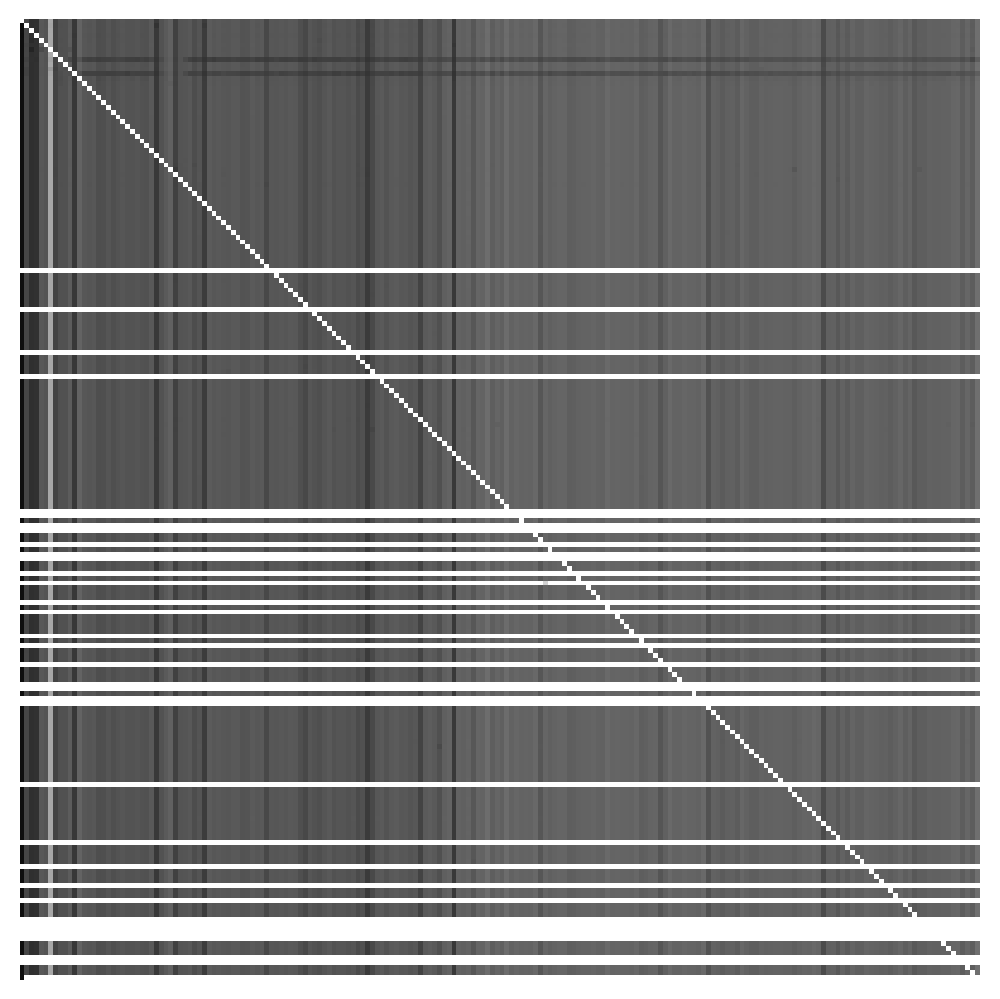}
\caption{$R$ obtained using $\boxtimes$ with $g(x)=\sqrt{x_{\rm b}}$}
\label{fig:boxtimesSQRTxb-case1}
\end{subfigure}
\caption{\it Positive evidence in the final referral trust matrix~$R$ for the Tribler data.
For each pair $(i,j)$ the amount of positive evidence underlying the opinion of
$i$ about $j$ is shown as a shade of gray, using a logarithmic gray scale.
White corresponds to zero, black to $8.7\cdot10^6$, which is the maximum amount of evidence
occurring in a single matrix entry in all the experiments.
}
\label{fig:referralR-case1}
\end{figure}

\begin{figure}[!t]
\centering
\begin{subfigure}[b]{.48\linewidth}
\includegraphics[width=60mm]{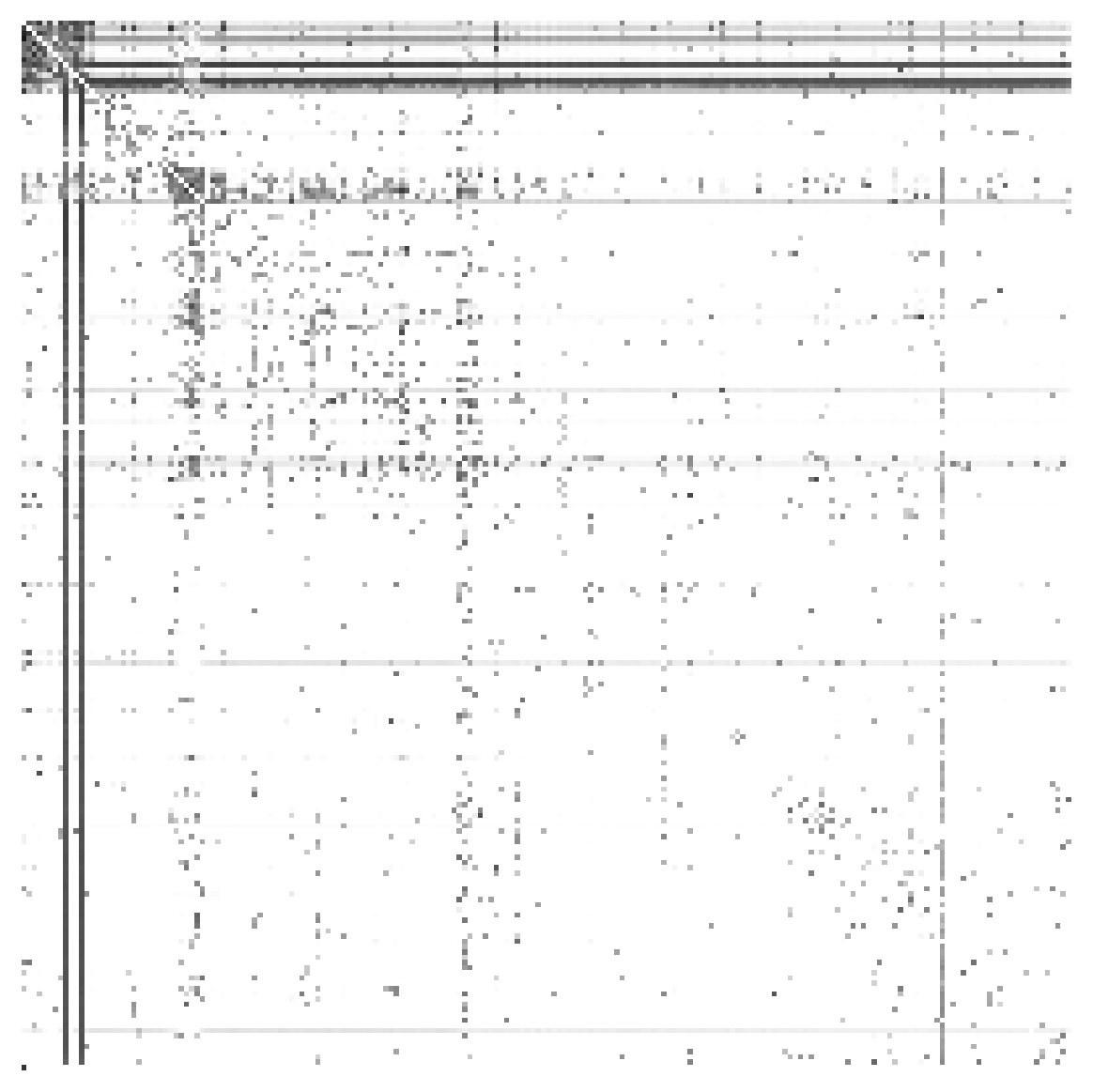}
\caption{$R$ obtained using $\odot$}
\label{fig:odot-case2}
\end{subfigure}
~~~
\begin{subfigure}[b]{.48\linewidth}
\includegraphics[width=60mm]{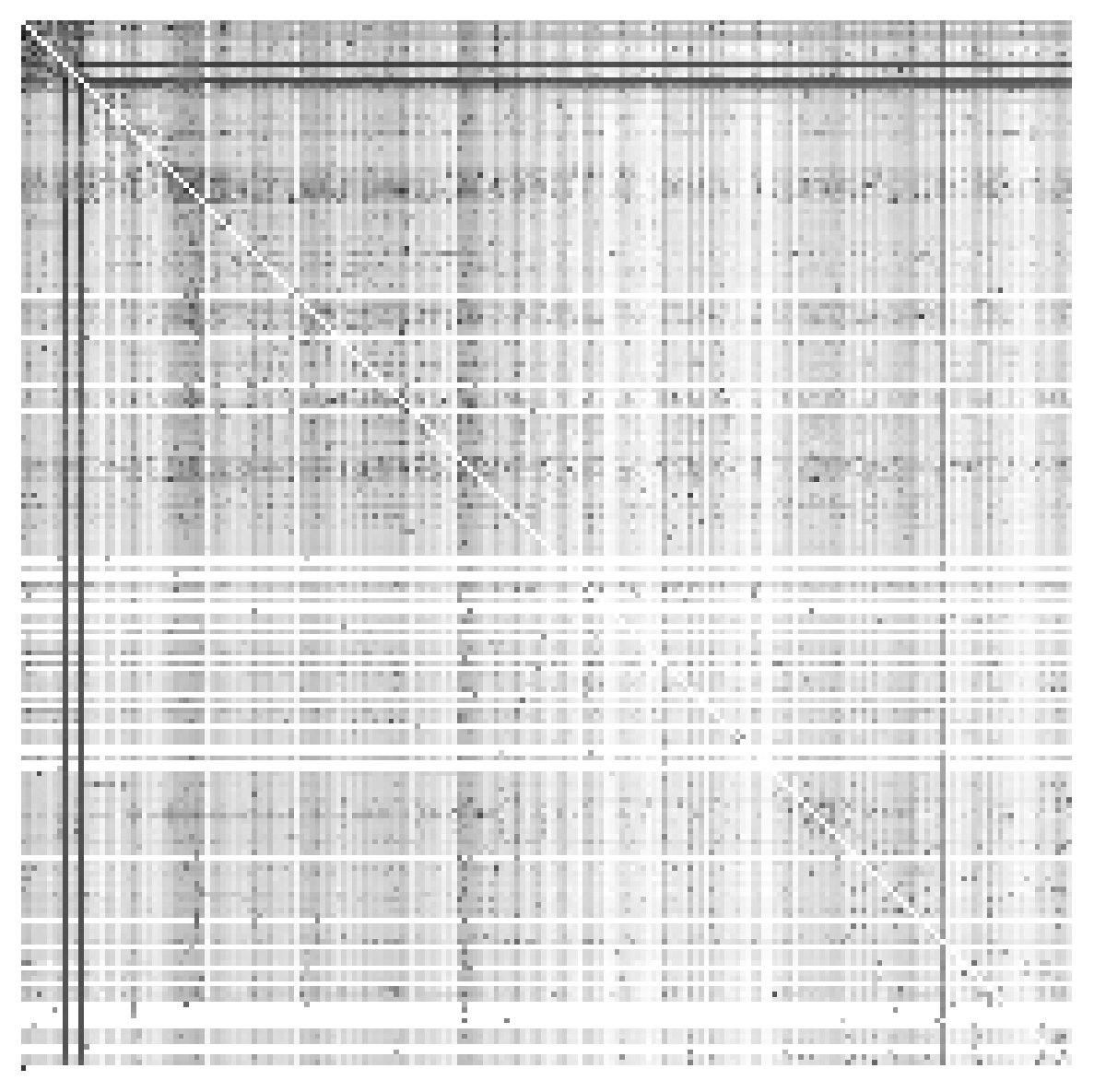}
\caption{$R$ obtained using $\otimes$}
\label{fig:otimes-case2}
\end{subfigure}
\begin{subfigure}[b]{.48\linewidth}
\includegraphics[width=60mm]{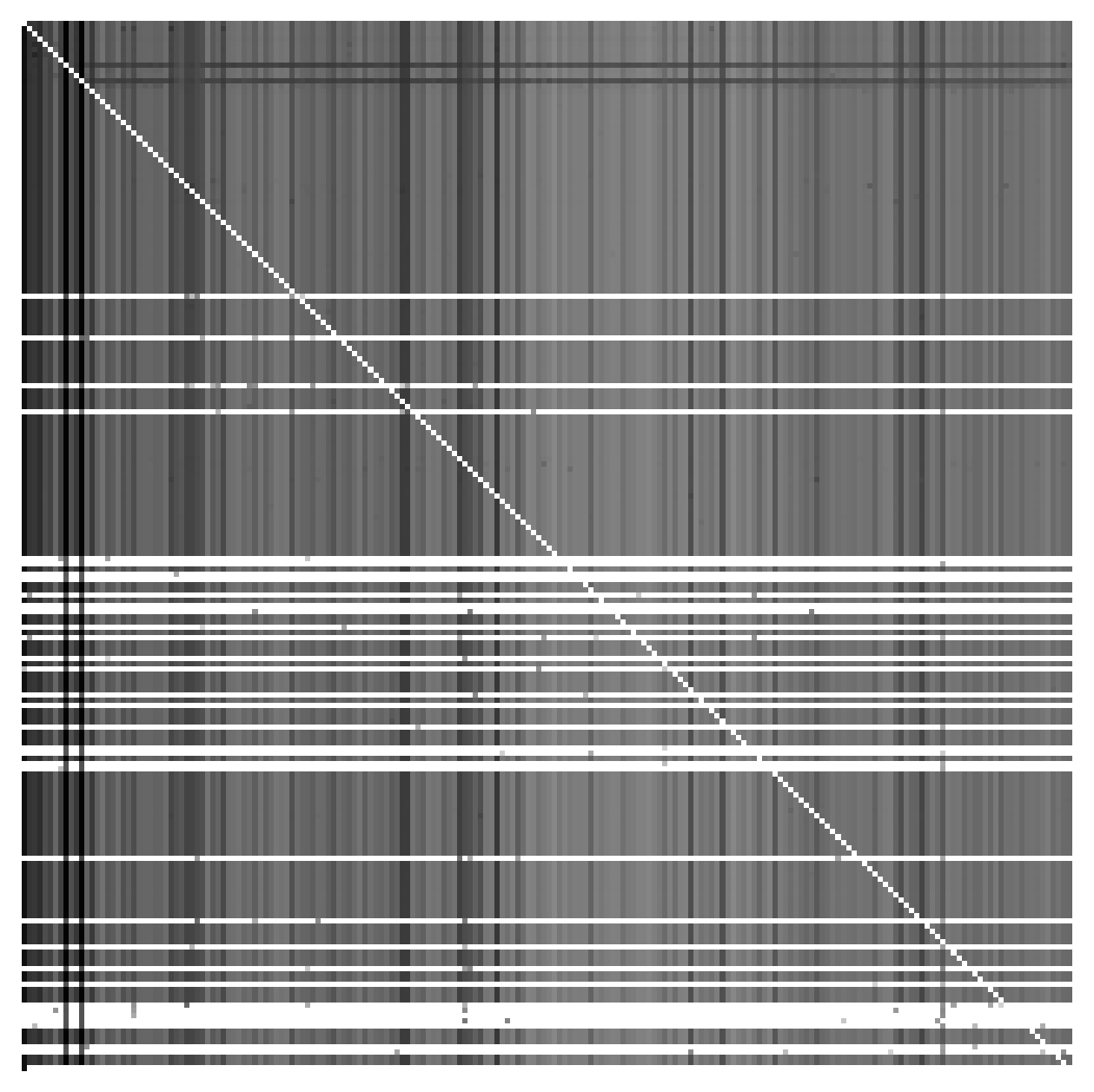}
\caption{$R$ obtained using $\boxtimes$ with $g(x)=x_{\rm b}$}
\label{fig:boxtimexb-case2}
\end{subfigure}
~~~
\begin{subfigure}[b]{.48\linewidth}
\includegraphics[width=60mm]{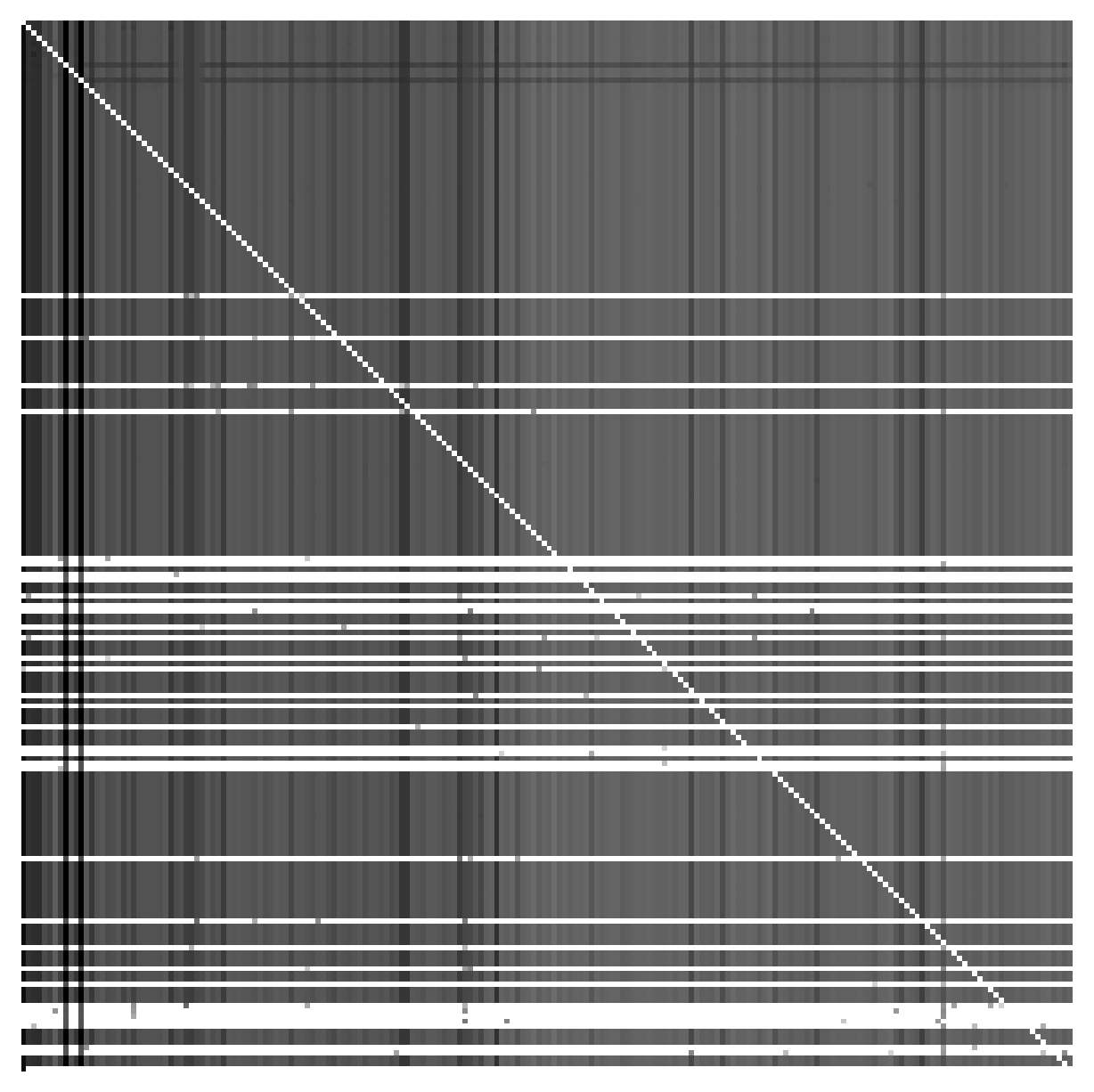}
\caption{$R$ obtained using $\boxtimes$ with $g(x)=\sqrt{x_{\rm b}}$}
\label{fig:boxtimesSQRTxb-case2}
\end{subfigure}
\caption{\it The sum of positive and negative evidence in the final referral trust matrix~$R$ for the Tribler data.
For each pair $(i,j)$ the total amount of evidence underlying the opinion of
$i$ about $j$ is shown as a shade of gray, using a logarithmic gray scale.
White corresponds to zero, black to $8.7\cdot10^6$, which is the maximum amount of evidence
occurring in a single matrix entry in the experiments.
}
\label{fig:referralR-case2}
\end{figure}

The amount of evidence in $R$ is presented in Figure~\ref{fig:referralR-case1} (only positive evidence) and in Figure~\ref{fig:referralR-case2} (sum of positive and negative evidence).
For most users, the amount of evidence in $R$ has increased 
compared to the initial situation (Figures~\ref{fig:initial-case1} and~\ref{fig:initial-case2}). 
The plots are characterized by uniform vertical stripes, indicating that (most) users have approximately the same amount of (positive) evidence about a given user.
The amount of evidence, however, remains close to $0$ for those users who had very few interactions with other users (horizontal white lines in Figures~\ref{fig:referralR-case1} and~\ref{fig:referralR-case2}). 
It is also worth noting that the diagonal of $R$ is clearly recognizable as a white line. 
This is due to the fact we impose the diagonal to be full uncertainty, i.e.\ users cannot have an opinion about themselves, to reduce the effect of self-promoting.

The choice of the discounting operator, which defines how evidence is propagated, has a significant impact on the amount of evidence in~$R$.
Ideally, users should be able to use the available trust information to decide whether to engage an interaction with another user~\cite{Vavilis2014}.
Therefore, a reputation system should allows users to gather as many recommendations (i.e., evidence) as possible from trusted users.
However, the use of the $\odot$ operator causes most of the evidence to be lost along the way.
This can be clearly understood by observing that the initial situation in Figure~\ref{fig:initial-case1} (Figure~\ref{fig:initial-case2} respectively) and the final referral trust matrix~$R$ in Figure~\ref{fig:odot-case1} (Figure~\ref{fig:odot-case2} respectively) are almost the same.
Figures~\ref{fig:otimes-case1} and~\ref{fig:otimes-case2} show that the $\otimes$ operator propagates more evidence than~$\odot$.
We remind the reader that $\otimes$ causes double-counting as well as discarding of evidence as shown in Examples~\ref{ex:strangelimit} and~\ref{ex:strangelimit2}.
The $\boxtimes$ operator with both $g(x)=x_{\rm b}$ and $g(x)=\sqrt{x_{\rm b}}$ results in the propagation of more evidence compared to the $\odot$ and $\otimes$ operators,
 as shown in Figures~\ref{fig:boxtimexb-case1} and~\ref{fig:boxtimesSQRTxb-case1} (positive evidence) and in Figures~\ref{fig:boxtimexb-case2} and~\ref{fig:boxtimesSQRTxb-case2} (total evidence).
These findings confirm the results obtained with the synthetic data (Table~\ref{tab:comparison-evidence}).
Therefore, we conclude that the $\boxtimes$ operator is preferable to the other operators.

\paragraph{Convergence}
We have analyzed the convergence of the iterative approach using the Tribler dataset.
The experiments  show that the reputation models built on top of EBSL  converge.
In particular, EBSL with $g(x)=x_{\rm b}$ converges after  47 iterations, 
EBSL with $g(x)=\sqrt x_{\rm b}$ converges after 24 iterations, and EBSL using the $\odot$ operator after 9 iterations. 
One can observe that, in all the cases, convergence is fast.
Accordingly, we believe that the proposed reputation model can handle real scenarios.
In the experiments we also analyzed the convergence of the na{\"i}ve approach that combines flow-based reputation and SL, as presented in Section~\ref{sec:flowSL}.
Here 
convergence is {\em not} reached in a reasonable amount of time: after 1000 iterations 
we still have $\sum_{i,j}\delta(R^{(k+1)}_{ij},R^{(k)}_{ij})\approx 10^{-8}$.


To study the link  between our approach and Markov chains, we performed additional experiments 
(not reported here) with a number of limiting situations, i.e. synthetic data unlikely to occur in real life.
In particular, we studied the EBSL case where the powers of $A$ show oscillations, i.e. 
$A^{k+2}=A^k$ with $A^{k+1}\neq A^k$.
Here $A^k$ stands for $((\cdots\boxtimes A)\boxtimes A)\boxtimes A$.
This can occur in Markov chains too.
In flow-based reputation (\ref{reputationrA}) the added term $(1-\alpha)s_x$ 
dampens the oscillations and thus improves convergence.
Similarly, in our EBSL experiments the added term $A$ in each iteration (\ref{eq:fix-point})
gives a convergent result in spite of the oscillatory nature of~$A$.
This strengthens our conviction that
EBSL correctly captures the idea of reputation flow.

 \section{Related Work}
\label{sec:related}

The notion of uncertainty is becoming an important concept in reputation systems and, more in general, in data fusion \cite{Bleiholder:2009,Keenan2011}.
Uncertainty has been proposed as a quantitative measure of the accuracy of predicted beliefs and it is used to represent the level of confidence in the fusion result.
Several approaches have extended reputation systems with the notion of uncertainty \cite{josang2001logic,ReganCP05,Ries11,Travos}.
For instance, Reis et al. \cite{Ries11} associate  a parameter with opinions to indicate the degree of certainty to which the average rating is assumed
to be representative for the future. 
Teacy et al. \cite{Travos} account for uncertainty by assessing the reputation of information sources based on the perceived accuracy of past opinions.
Differently from the previous approaches, Subjective Logic \cite{josang2001logic} 
considers uncertainty as a dimension orthogonal to belief and disbelief, 
which is based on the amount of available evidence.

One of the main challenges in reputation systems is how to aggregate opinions, 
especially in the presence of uncertainty.
SL provides two main operators for aggregating opinions: consensus and discounting (see Section~\ref{sec:prelimsubjective}).
Many studies have analyzed strategies for combining conflicting beliefs 
\cite{Josang2002157,Josang03,Smets2007387} 
and have proposed new combining strategies and operators \cite{Cerutti:2013,Zhou2011}. 
In Section~\ref{sec:evidencemap}, we re-confirm that the standard consensus operator used in SL 
is well-founded on the theory of evidence.

In contrast, less effort has been devoted to studying the discounting operator.
Bhuiyan and J{\o}sang \cite{Bhuiyan10} propose two alternative discounting operators: an operator based on opposite belief favouring, for which the combination of two disbeliefs results in belief, and a base rate sensitive transitivity operator in which the trust in the recommender is a function of the base rate.
Similarly to the traditional discounting operator of SL, 
these operators are founded on probability theory.
As shown in Section~\ref{sec:flowSL}, employing operators founded on different theories has the disadvantage that these operators may not ``cooperate''.
In the case of SL, this lack of cooperation results in 
the inability to apply SL to 
arbitrary trust networks.
In particular, trust networks have to be expressed in a canonical form in which edges are not repeated. 
A possible strategy to reduce a trust network to a canonical form 
is to remove the weakest edges (i.e., the least certain paths) until the network can be expressed in canonical form \cite{Josang2006simplification}. 
This, however, has the disadvantage that some (possibly even much) trust information is discarded.
An alternative canonicalization method called edge splitting was presented in \cite{Josang08}.
The basic idea of this method is to split a dependent edge into a number of different edges 
equal to the number of different instances of the edge in the network expression.
Nonetheless, the method requires that the trust network is acyclic; if a loop occurs in the trust network, some edges have to be removed in order to eliminate the loop, thus discarding trust information.
In contrast, we have constructed a discounting operator founded on the theory of evidence.
This operator together with the consensus operator allows the computation of reputation for arbitrary trust networks, which can include loops, without the need to discard any information.

Cerutti et al. \cite{Cerutti:2013} define three requirements for discounting based on 
the intuitive understanding of few scenarios: 
Let $A$ be $x$'s opinion about $y$'s trustworthiness, 
$C$ the level of certainty that $y$ has about a proposition $P$, 
and $F= A \circ C$ the (indirect) opinion that $x$ has about~$P$. 
(i) If $C$ is pure belief, then $F=A$; 
(ii) If $C$ is complete uncertainty, then $F=C$; 
(iii) The belief part of $F$ is always less than or equal to the belief part of~$A$.
Based on these requirements, they propose a family of graphical discounting operators which, 
given two opinions, project one opinion into the admissible space of opinions given by the other opinion.
These operators are founded on geometric properties of the opinion space.
This makes it difficult to determine whether the resulting theory is consistent with the theory 
of evidence or probability theory.
Our discounting operator satisfies requirement~(ii) above, but not requirements (i) and (iii); 
indeed, for $g(A)>0$ it holds that $A\boxtimes B = B$ (where $B$ represents full belief).
It is worth noting that the requirements proposed in \cite{Cerutti:2013} 
are not well founded in the theory of evidence: 
$B$ means that there is an infinite amount of positive evidence; 
discounting an infinite amount of evidence still gives an infinite amount of evidence.
In Theorem~\ref{th:simplemap}, we provided a number of desirable properties founded on the theory of evidence.
In particular, if $p+n\rightarrow \infty$ then $u\rightarrow 0$.
Accordingly, if $C=B$ the uncertainty component of $F$ should be equal to 0, 
regardless of the precise (nonzero) value of the uncertainty component of $A$.


To our knowledge, our proposal is the first work that integrates uncertainty into flow-based reputation.


\section{Conclusion}
\label{sec:conclusions}

In this paper, we have presented a flow-based reputation model with uncertainty that allows the construction of an automated reputation assessment procedure for arbitrary trust networks.
We illustrated and discussed the limitations of a na{\"i}ve approach to combine flow-based reputation and SL.
An analysis of SL shows that the problem is rooted in the lack of ``cooperation'' between the SL consensus and discounting rules due to the different nature of these two operators.
In order to solve this problem, we have revised SL by introducing a scalar multiplication operator 
and a new discounting rule based on the flow of evidence.
We refer to the new opinion algebra as Evidence-Based Subjective Logic (EBSL).

A generic definition of discounting (the operator $\boxtimes$) lacks the associative property satisfied 
by the SL operator $\otimes$.
This, however, is not problematic since the flow of evidence has a well defined direction.
Furthermore, the operator $\boxtimes$ has right-distributivity, a property that one would intuitively expect
of opinion discounting.  
One can choose a specific discounting function $g(x)$ proportional to the amount of positive evidence in~$x$.
The resulting discounting operator is denoted as $\odot$.
As shown in Table~\ref{tab:properties}, this operator is completely linear (associative as well as
left and right distributive).
However, it has potentially undesirable behavior since it ignores negative evidence,
and requires a carefully chosen system parameter related to the maximum amount 
of positive evidence in the system.

The adoption of the discounting operator $\boxtimes$ results in a system that is centered entirely on the handling of evidence. 
We have showed that this new EBSL algebra makes it possible to define an iterative algorithm to compute reputation for arbitrary trust networks.
Thus, EBSL poses the basis for the development of novel reputation systems.
In particular, our opinion algebra guarantees that trust information does not have to be discarded, 
thus preserving the quality of the aggregated evidence.
Moreover, making the uncertainty of the computed reputation explicit helps users in deciding how much to rely on it based on their risk attitude.
In our opinion, this will facilitate the adoption and acceptance of reputation systems since users are more aware of the risks of engaging a transaction and, thus, can make more informed decisions.

\begin{table}[!t]
\centering
{\small
\begin{tabular}{|l|c|c|c|}
\hline
& $\otimes$ & $\boxtimes$ & $\odot$ \\
\hline
Associativity &yes& no& yes\\
\hline
Left-distribution &no & no & yes \\
\hline
Right-distribution &no&yes&yes\\
\hline
Recursive solutions &no&yes&yes\\
\hline
\end{tabular}
}
\caption{\it Comparison of the operators $\otimes$, $\boxtimes$, and $\odot$.}
\label{tab:properties}
\end{table}




The work presented in the paper poses the basis for several directions of future work.
We have shown how EBSL can be used to build a flow-based reputation model with uncertainty.
However, several reputation models have been proposed in the literature to compute reputation over a trust network.
An interesting direction is to study the applicability of EBSL as a mathematical foundation for these models.
This will also make it possible to study the impact of uncertainty on the robustness of reputation systems against attacks like self-promotion, slandering and Sybil attacks.


\begin{acknowledgements}
This work  has been partially funded by the Dutch national program COMMIT under the THeCS project, the ITEA2 project FedSS, 
the ARTEMIS project ACCUS, and the EDA project IN4STARS~2.0.
\end{acknowledgements}

%
%
\bibliographystyle{spmpsci}
\bibliography{reputation}


\appendix

\section{Associativity and Left-Distributivity of $\boxtimes$ Imply Linearity of $g$}

We present a proof of Lemma~\ref{lemma:linearg}.
The proof technique requires that opinions and the operations
$\oplus$ and $\cdot$
form a vector space with an inner product defined on it.
There is no such thing as an opinion `$-x$' 
(inverse of $x$ with respect to $\oplus$) 
in $\Omega$ or $\Omega'$,
so we have to broaden our scope.
We define an extended opinion space
\[
	\Omega^* \stackrel{\rm def}{=}\{(b,d,u)\in[-1,1]^2\times[0,1]\;\;|\;\;
	|b|+|d|+u=1\} 
\]
that allows for negative belief and/or disbelief components.
We have to slightly modify the relation between evidence and opinions,
so that negative {\em amounts} of evidence can be represented,
\begin{eqnarray*}
	(b,d,u)=\frac{(p,n,c)}{|p|+|n|+c} 
	\quad&;&\quad
	(p,n)=c\frac{(b,d)}{u}
\end{eqnarray*}
with $c>0$. Here $p$ and $n$ can be negative.
This relation automatically leads to a slightly modified definition
of evidence addition ($\oplus$) and scalar multiplication,
\begin{eqnarray*}
	x\oplus y &\stackrel{\rm def}{=}& 
	\frac{(\xb\yu+\yb\xu,\; \xd\yu+\yd\xu,\; \xu\yu)}{|\xb\yu+\yb\xu|+|\xd\yu+\yd\xu|+\xu\yu}
	\\
	\alpha\cdot x &\stackrel{\rm def}{=}& 
	\frac{(\alpha\xb,\; \alpha\xd,\; \xu)}{|\alpha\xb|+|\alpha\xd|+\xu}.
\end{eqnarray*}
For $x,y\in\Omega$ and $\alpha\geq 0$ all this reduces to the algebra
of Sections \ref{sec:prelimsubjective} and~\ref{sec:scalar};
For $x,y\in\Omega^*$ and $\alpha\in{\mathbb R}$ all the nice linear properties
still hold.

The space $\Omega^*$ with the $\oplus$ and $\cdot$ operations is a vector space.
(The underlying space of $(p,n)$ evidence pairs has also been turned into a vector space
by allowing negative amounts of evidence.)
We introduce an inner product on this vector space as follows,
\[
	\langle x,y\rangle \stackrel{\rm def}{=} p(x)p(y)+n(x)n(y).
\]
It is easily verified from the definitions that this expression 
satisfies all the requirements for being an inner product, namely
$\langle x,y\rangle=\langle y,x\rangle$;
$\langle x,y\oplus z\rangle=\langle x,y\rangle+\langle x,z\rangle$;
$\langle x,x\rangle\geq 0$ and
$\langle \alpha\cdot x,y\rangle=\alpha \langle x,y\rangle$.

With all this structure in place we can now invoke the 
Riesz-Fr\'echet theorem \cite{Heuser2006}, 
\begin{quote}
	If $\Omega^*$ is a real Hilbert space and $g$ a linear functional, 
	then there exist a unique vector $v\in \Omega^*$ 
	such that $g(x)=\langle v, x\rangle$ for all $x\in \Omega^*$.
\end{quote}
Here `linear functional' means that the linear property $g(x\oplus y)=g(x)+g(y)$ holds.
Hence, the only way to achieve this property is to set 
$g(x)=\langle v,x\rangle$, i.e. a linear combination of $p(x)$ and $n(x)$.

\end{document}